\documentclass[a4paper,twoside]{article}
\newcommand{\affil}[1]{$^{\rm #1}$}

\usepackage[authoryear]{natbib}
\bibpunct{(}{)}{;}{a}{}{,}
\usepackage{graphicx}
\usepackage{xspace}
\usepackage{color}
\usepackage{stabular}
\date{} 
\def\plateOne{364\xspace}
\def\plateZero{357\xspace}
\def\redshifts{400\xspace}

\def\zmid{{\boldmath $z \sim 0.77$}\xspace}
\def\zm{$z \sim 0.77$\xspace}

\def\farcs{\hbox{$.\!\!^{\prime\prime}$}}
\def\arcmin{\hbox{$^\prime$}}
\def\arcsec{\hbox{$^{\prime\prime}$}}
\title{\large\bf\flushleft An Efficient Approach to Obtaining Large Numbers
  of Distant Supernova Host Galaxy Redshifts}
\author{\parbox{\textwidth}{\flushleft
\vspace{-0.5cm}
{\it C. Lidman\affil{\,A,B,Q}, V. Ruhlmann-Kleider\affil{\,C},
  M. Sullivan\affil{\,D}, J. Myzska\affil{\,A,E},
  P. Dobbie\affil{\,A}, K. Glazebrook\affil{\,F,B},
  J. Mould\affil{\,F,B}, P. Astier\affil{\,G},C. Balland\affil{\,G,P},
  M. Betoule\affil{\,G}, R. Carlberg\affil{\,H}, A. Conley\affil{\,I}, D. Fouchez\affil{\,J}, J. Guy\affil{\,G}, D. Hardin\affil{\,G}, I. Hook\affil{\,D,N}, D. A. Howell\affil{\,K,O}, R. Pain\affil{\,B,G}, N. Palanque-Delabrouille\affil{\,C}, K. Perrett\affil{\,H,L}, C. Pritchet\affil{\,M}, N. Regnault\affil{\,G}, and J. Rich\affil{\,C}}\\
\vspace{0.4cm}
{\small \affil{A}\,Australian Astronomical Observatory, PO Box 296, Epping NSW 1710, Australia}\\
{\small \affil{B}\,ARC Centre of Excellence for All-sky Astrophysics (CAASTRO)}\\
{\small \affil{C}\,DSM/IRFU/SPP, CEA-Saclay, F-91191 Gif-sur-Yvette, France}\\
{\small \affil{D}\,Department of Physics (Astrophysics), University of Oxford, DWB, Keble Road, Oxford OX1 3RH, UK}\\
{\small \affil{E}\,Villanova University, 800 East Lancaster Avenue  Villanova, PA 19085, USA}\\
{\small \affil{F}\,Centre for Astrophysics and Supercomputing, Swinburne University of Technology, PO Box 218, Hawthorn, VIC 3122, Australia}\\
{\small \affil{G}\,LPNHE, Universit´e Pierre et Marie Curie Paris 6, Universit´e Paris Diderot Paris 7, CNRS-IN2P3, 4 place Jussieu, 75005 Paris, France}\\
{\small \affil{H}\,Department of Astronomy and Astrophysics, University of Toronto, 50 St. George Street, Toronto, ON, M5S 3H4, Canada}\\
{\small \affil{I}\,Center for Astrophysics and Space Astronomy, University of Colorado, 593 UCB, Boulder, CO, 80309-0593, USA}\\
{\small \affil{J}\,CPPM, CNRS-IN2P3 and University Aix Marseille II, Case 907, 13288 Marseille cedex 9, France}\\
{\small \affil{K}\,Las Cumbres Observatory Global Telescope Network, 6740 Cortona Dr., Suite 102, Goleta, CA 93117, USA}\\
{\small \affil{L}\,DRDC Ottawa, 3701 Carling Avenue, Ottawa, ON, K1A 0Z4, Canada}\\
{\small \affil{M}\,Department of Physics \& Astronomy, University of Victoria, PO Box 3055, Stn CSC, Victoria, BC, V8W 3P6,  Canada}\\
{\small \affil{N}\,INAF, Osservatorio Astronomico di Roma, via Frascati 33, 00040 Monteporzio (RM), Italy}\\
{\small \affil{O}\,Department of Physics, University of California, Santa Barbara, Broida Hall, Mail Code 9530, Santa Barbara, CA 93106-9530, USA}\\
{\small \affil{P}\ Universit´e Paris 11, Orsay, F-91405, France}\\
{\small \affil{Q}}\,Email: clidman@aao.gov.au}}
\begin{document}
\twocolumn[
\begin{changemargin}{.8cm}{.5cm}
\begin{minipage}{.9\textwidth}
\vspace{-1cm}
\maketitle
%
%
\small{\bf Abstract: - We use the wide--field capabilities of the 2dF
  fibre positioner and the AAOmega spectrograph on the
  Anglo-Australian Telescope (AAT) to obtain redshifts of galaxies
  that hosted supernovae during the first three years of the Supernova
  Legacy Survey (SNLS). With exposure times ranging from 10 to 60 ksec
  per galaxy, we were able to obtain redshifts for \redshifts host
  galaxies in two SNLS fields, thereby substantially 
  increasing the total number of SNLS supernovae with host galaxy redshifts. The median redshift of the galaxies in
  our sample that hosted photometrically classified Type Ia supernovae
  (SNe Ia) is \zmid, which is 25\% higher than the median redshift of
  spectroscopically confirmed SNe Ia in the three--year sample of the
  SNLS. Our results demonstrate that one can use wide--field fibre--fed
  multi--object spectrographs  on 4m telescopes  to efficiently obtain redshifts for large
  numbers of supernova host galaxies over the large areas of sky that will
  be covered by future high--redshift supernova surveys, such as the
    Dark Energy Survey.}


\medskip
\medskip
\end{minipage}
\end{changemargin}
]
\small

\section{Introduction}

The landmark discovery of the accelerating Universe more than a dozen
years ago \citep{Perlmutter1999, Riess1998} was made from observations
of a few dozen distant Type Ia supernovae (SNe Ia). Since then, SN Ia
samples have grown in size by an order of magnitude. Currently, the
largest published individual sample of distant SNe Ia is the 3--year sample from
the Supernova Legacy Survey (SNLS), which contains 252 SNe Ia
extending out $z=1.06$ \citep{Guy2010}. The Dark Energy Survey (DES),
which will discover $\sim$4\,000 distant SNe Ia \citep{Bernstein2011},
is yet another order of magnitude larger.

All of the SNe Ia in the 3--year SNLS sample were spectroscopically
confirmed.  To confirm a SN Ia at $z\sim 0.7$ from its spectral
features (and to obtain its redshift) typically takes about an hour
 of exposure time with an efficient spectrograph on an
8--10\,m class telescope. These SNe are observed one at a time, as
it is, at any instant, very rare to have more than one SN
Ia within one week of maximum light visible within the fields of view
that are available with the current generation of spectrographs on
8--10\,m class telescopes. Assuming 100\% efficiency, the Dark Energy
Survey will need around 4\,000 hours of 8--10\,m class telescope time
 with current instrumentation to spectroscopically
confirm the 4\,000 SNe Ia that they will find. Clearly, this
 constitutes a very large request, so alternative
  approaches should be investigated first.

A promising yet unproven alternative is to classify SNe Ia using
multi--colour lightcurves and to obtain the redshifts from host
galaxies after the SNe have faded from view. The principle advantages
of this alternative are that i) host galaxy redshifts can be
 obtained at any time, and  ii) one can
  observe several hosts simultaneously if the number density of host
  galaxies is sufficiently high enough.
 
In this paper, we describe how we used the 2dF fibre positioner and the
AAOmega spectrograph on the Anglo--Australia Telescope (AAT) to obtain
redshifts of galaxies that hosted supernovae in two of the four SNLS
fields. A second paper reporting the results from a 3rd SNLS field
(the fourth field is not visible from the AAT)  and the
  analysis of the three fields combined will be written once
observations of that field have been taken. The current paper is
divided as follows. In Section \ref{Sec:TargetSelection}, we briefly
describe the characteristics of the SNLS and explain how targets from
the SNLS were selected and prioritised for spectroscopy. Then, in
Section \ref{Sec:Observations} and Section \ref{Sec:DataReduction}, we
describe the observations and outline the steps used to process the
data. In the following two sections, we describe our results
(Section~\ref{Sec:Results}) and discuss (Section~\ref{Sec:Discussion})
how our observing strategy compares to an observing strategy in which
transients are observed in real time. In the final section, we
summarise our main conclusions.  Unless noted otherwise, we use AB
magnitudes throughout this paper.

\section{SNLS Host Galaxies}\label{Sec:TargetSelection}

The SNLS consisted of a rolling search of four one--square degree
fields, labeled D1 to D4, using the MegaCam camera
\citep{Boulade2003} on the Canada--France--Hawaii Telescope
(CFHT)\footnote{The SNLS fields were common to the deep synoptic
  survey of the Canada--France--Hawaii Telescope Legacy Survey --
  http://www.cfht.hawaii.edu/Science/CFHLS/}.  During dark and grey
time and over a period of of 5 to 7 consecutive lunations, the fields
were imaged every 3 to 4 nights in $g_{\mathrm{M}}$, $r_{\mathrm{M}}$,
$i_{\mathrm{M}}$, and $z_{\mathrm{M}}$   filters. The survey ran for five years,
starting in 2003 and ending in 2008. Additional details on the survey
can be found in \citet{Astier2006, Sullivan2006a} and
\citet{Perrett2010}.

During the course of the SNLS, transient events were detected,
classified and prioritised in real time
\citep{Perrett2010,Sullivan2006a}. The highest priority SN Ia
candidates were then sent to telescopes at the Keck, ESO Paranal and
Gemini Observatories for spectroscopic typing and redshifting. The
spectroscopic follow--up of real--time transients are described in
\citet{Howell2005}, \citet{Bronder2008}, \citet{Ellis2008},
\citet{Balland2009}, \citet{Walker2010} and Fahkouri et al. (in
preparation).

Likely SNe Ia were prioritised according to several criteria, such as,
among others, the quality of the light curve fit, the measured
stretch, peak apparent brightness and light curve phase, and the ratio
of the likely peak flux of the SN Ia with respect to the flux of the
host \citep{Perrett2010}. For example, SNe Ia that were likely to have
a peak observer--frame i--band magnitude fainter than
$i_{\mathrm{AB}}=24.4$ were not sent for spectroscopy, since these SNe
were likely to be beyond $z=0.9$, are difficult to confirm
spectroscopically, and generally have one well--measured colour at most. The selection
criteria are not treated in isolation. For example, candidates in the
magnitude range $23.8 < i_{\mathrm{AB}} < 24.4$ were only followed if
the flux of the candidate was as large as the host. Due to the
vagaries of telescope scheduling and weather, not all high priority
targets could be observed, and on some classically scheduled nights,
  lower priority targets were followed once the higher priority targets
had been observed.

The spectroscopically confirmed sample is about 60\% complete up to
$z\sim 0.6$ \citep{Perrett2010}.  At this redshift, the incompleteness
is mostly due to the selection criteria that are applied to the lightcurve
data. Completeness then falls steadily as redshift increases. By
$z\sim 1$, the completeness drops to 20\% and then to zero by $z\sim
1.2$. The most distant spectroscopically confirmed SNe Ia in the
3--year SNLS sample is at $z=1.06$ \citep{Guy2010}.

 Therefore there are as many SNe Ia detected in the SNLS that lack
spectroscopy as there are SNe Ia that were spectroscopically
confirmed. Furthermore, most of the untargeted SNe Ia will be at
higher redshifts. These SNe Ia are now far too faint to observe
spectroscopically, so they can no longer be classified in this way;
however, with full light curves, many of them can be classified
photometrically \citep{Bazin2011}. Together with a precise measure of
the redshift, which can be obtained from follow--up spectroscopy of the
host, these SNe Ia can be added to the Hubble diagram and used to
constrain the expansion history of the Universe, provided that the
light curves are of sufficient quality and only after biases specific
to both the photometric selection of the SNe Ia and the spectroscopic
follow--up of the hosts are modelled and correctly accounted for.

Besides SNe Ia, the SNLS also contains a couple of hundred core--collapse
supernovae (CC SNe) below $z=0.4$. Since CC SNe were rarely followed
spectroscopically, only a fraction of these (42 in the first three
years of SNLS) were spectroscopically confirmed. As shown in
\citet{Bazin2009}, photometric classification allows one to identify
CC SNe and, together with a measure of the redshift, the CC SNe
rate. Precise host galaxy redshifts can reduce the uncertainty of this
measurement significantly.


\subsection{Target Selection}\label{Sec:TargetSelection}

\citet{Bazin2011} reanalysed the first three years of SNLS survey and
selected candidate supernovae based on the full three year light
curves. Their catalogue contains 1483 candidates, of which 1233 were
matched to a galaxy with a reliable photometric redshift (see
\citet{Bazin2011} for details). Of these 1233 candidates, 485 were
photometrically classified as SN Ia.  In total, 176 of the
  485 photometrically classified SN Ia were confirmed
  spectroscopically. None of the photometrically classified SNe Ia
  were confirmed to be supernovae of other types or AGNs, however,
  not all 485 candidates were targeted for spectroscopy during the
  real--time follow--up.

For the follow--up spectroscopy with AAOmega, we split the catalogue
of 1483 candidate supernovae into five categories and assign a
priority to each category. Priority 5  --- the highest
  priority we assign to photometrically identified SNe --- goes to
photometrically classified SNe Ia that lack a spectroscopic
redshift. Then, in order of decreasing priority were photometrically
classified CC SNe without a spectroscopic redshift, candidate
supernovae of unknown type with an assigned host but without a
spectroscopic redshift (most objects in this priority have a reliable
photometric redshift), candidate supernovae of unknown type without an
assigned host and without a redshift of any kind (photometric or
spectroscopic), and finally, the lowest priority (priority 1),
candidate supernovae of any kind with a spectroscopic redshift. If the
identity of the host was unclear, the fibre was placed at the location
of the supernova. Otherwise, it was placed on the nearest host.

\begin{table*}[h]
\begin{center}
\caption{Target selection summary}\label{table:targetSelection}  
\vspace{0.1cm}
\begin{tabular}{llcrrr}
\hline\hline
Field & Target Type $^a$              & Priority $^b$    & Number of         & Number & Completeness \\
        &                                   &        & potential targets	  & targeted          & fraction$^c$\\ 
\hline                           
D1    & White dwarfs                       &     6             & 7       & 7      & ...     \\       
D1    & SNe Ia without $z_{s}$         & 5             & 70      & 69     & 45  \\
D1    & CC without $z_{s}$              & 4             & 18      & 17     & 94  \\
D1    & SNe with a host but no $z_{s}$           & 3             & 218     & 201    & 52  \\
D1    & SNe without hosts and $z$        & 2             & 21      & 18     &   6  \\
D1    & SNe with $z_{s}$                  & 1             & 117     & 49     & 53  \\
D1    & Cluster galaxies                  & 1--5          & 13      & 7      & 54  \\
&&&&&\\
\hline 
D4    & White dwarfs                      & 6             & 7       & 7      & ...     \\
D4    & SNe Ia without $z_{s}$        & 5             & 65      & 64     & 77  \\
D4    & CC without $z_{s}$             & 4             & 11      & 10     & 60  \\
D4    & SNe with a host but no $z_{s}$           & 3             & 129     & 125    & 60  \\
D4    & SNe without hosts and $z$      & 2             & 32      & 30     &  13  \\
D4    & SNe with $z_{s}$                 & 1             & 125     & 115    & 76  \\
&&&&&\\
\hline 
D1 \& D4   & All SNe             & 1--5           & 806     & 698    & 57 \\
\hline
\end{tabular}
\end{center} 
\medskip

$^a$ $z_{s}$ represents the spectroscopic redshift: $z$ is a
redshift of any kind,  i.e.~spectroscopic or photometric.\\
$^b$ The priority goes from 6 (highest) to 1 (lowest).\\
$^c$ The completeness fraction is the number of objects with
a redshift quality flag of 3,4 or 5 divided by the number of objects
that were targeted with AAOmega.\\ 
\end{table*}

In addition to SN hosts, we targeted white dwarfs from
\citet{Limboz2008} and from our own search of the SNLS fields, and in
the 2hr field only, galaxies in X--ray selected clusters from the
XMM--LSS survey \citep{Pierre2006}. Both kinds of targets are
interesting for a number of reasons related to cosmology. Hot DA white
dwarfs are often used as flux calibrators. If a sufficient number of
hot DA white dwarfs can be found within the supernova search fields,
then an alternative and perhaps more precise method of calibrating SNe
Ia fluxes would become available.  These objects were
  given the highest priority (priority six) as we wanted to make sure
  that they were all allocated a fibre by the automated fibre
  allocation software \citep{Miszalski2006}. At the other end of the
mass scale, the number density and clustering of galaxy clusters and
the evolution of these properties with redshift depend on the
properties of dark energy and can therefore be used to constrain
it. The number of targets in each category and their priorities are
listed in Table~\ref{table:targetSelection}.

\subsection{Fibre allocation}

The 2dF fibre positioner positions fibres over a 2-degree
  diameter field. The absolute minimum separation between fibres is
  30\arcsec\ and is set by the rectangular shape of the magnets that
  hold the fibre buttons to the metallic field plate. If the
  separation between a given pair of supernova hosts is smaller than
  the minimum permitted fibre separation, then both hosts cannot be
  observed at the same time. The allocation of fibres to targets is
  then done on the basis of this requirement, the number of available
  fibres and target priorities \citep{Miszalski2006}.

From three years of SNLS data, the number of supernovae per SNLS field
is slightly larger than the number of fibres that are available with
2dF.  Each SNLS field has a field-of-view of 1 sq. degree,
  so there is the additional constraint that the fibres need to be
  placed within an area that is about a factor of 3 smaller than the
  full field available with 2dF. It is interesting to see what
fraction of the fibres were allocated to supernova hosts. A fraction
that is significantly smaller than 100\% is an indication of
significant clustering.

As an example, we compute the fraction of fibres that were assigned to
supernova host galaxies for the D4 field, which has the smallest
difference between the number of targets and the number of fibres. The
numbers listed in Table~\ref{table:targetSelection} cannot be used to
compute this fraction directly because multiple configurations were
used for the D4 field. Instead we simply examine the fraction of
fibres that were assigned in each configuration individually and
average the results.

For the D4 field, there were 362 supernovae of all priorities. Out of
339 fibres available for plate one and 332 fibres available for plate
zero\footnote{At the time of the observations, \plateZero fibres were
  available on plate 0 and \plateOne fibres were available on plate
  1. After subtracting 25 fibres that were used to measure the sky
  background approximately 330 fibres were used to select targets in
  the SNLS fields.}, respectively, 332
and 325 of them were allocated to supernova hosts.  The
remaining fibres were allocated to white dwarfs, which had higher
priority than the supernova hosts. The allocation rate of fibres to
supernova hosts was above 95\% and would have been 100\% if only the
supernova hosts were targeted.

Interestingly, \citet{Carlberg2008} finds that SNe Ia are
clustered more strongly than the average galaxy. A similar study for
CC SNe has not been done.  While there certainly were cases where two host galaxies
could not be observed simultaneously because they were too close to
each other, the supernova catalogue was large enough to ensure that
all fibres could be used efficiently.

The field covered by the DES camera \citep{Flaugher2010} is three times larger than MegaCam
on CFHT, the camera used by SNLS, and is similar to the field covered
by 2dF. If the number of supernova discoveries per season and per unit
area is similar for DES and SNLS, then it would take only one year
of searching with the DES camera to find as many supernova per
pointing as there are 2dF fibres. These supernova will be spread over
the entire 2dF field, so the chance of fibre collisions will be smaller
than it was for the follow--up of SNLS host galaxies.

\section{Observations}\label{Sec:Observations}

Fibres from 2dF feed AAOmega \citep{Smith2004}, which is a
bench--mounted double--beam spectrograph that is located in one of the
Coud\'{e} rooms at the base of the AAT. The red and blue arms of the
spectrograph are split by a dichroic. In the blue arm, we used the
580V grating: in the red arm, we used the 385R grating. The combined spectra
cover the wavelength interval from 3300\,$\AA$ to
8840\,$\AA$ at  spectral resolution of about $1400$.

The data were taken during clear conditions. The image
quality, as measured by the 2df focal plane imager, which can be
positioned in front of the observing plate but behind the 2df
corrector, varied between 1\farcs1 and 1\farcs7. Each fibre has a
diameter of approximately 2\farcs1. Distortions from the corrector
mean that fibres near the field edge subtend a slightly larger region
than fibres in the centre. The data were collected over four
consecutive nights starting on August 24th, 2011.

Each integration lasted 2500 seconds and a given configuration was
observed between two to four times before switching to a new
configuration. Immediately after a new configuration was tumbled into
place, a fibre flat--field and a fibre arc were obtained. 

Each SNLS field was configured  for plate 0 and again for
plate 1. Since the number of fibres for each plate are different, not
all objects observed in plate 0 were also observed in plate 1 and
visa--versa. Half way through the run, we processed the data that had
been taken for the D4 field by that time, allowing us to
re--prioritise objects that could be  assigned a redshift
  from clearly identified spectral features.

All  data were taken with the telescope pointing above an
altitude of 40 degrees. The total
integration times for the D1 and D4 fields  were 32,500 and 60,000 seconds,
respectively.  The time spent on any given target, however,
might have been less than this as it may have not have been allocated
a fibre in all the configurations. Less time was
spent on the D1 field because of time lost to cloud during the 2nd half of the
3rd night and because of the smaller number of hours the D1  field was at a
suitable hour angle. At the end of each night, five 2500 second dark
frames were taken.

In order to derive an estimate of the sky background,
  which is used in the processing of the data, 25 fibres were
allocated to observe regions within the SNLS fields that were free of
objects. We used version T0006 of the stacks
produced by TERAPIX\footnote{http://terapix.iap.fr/} to select these
regions. These stacks reach a limiting magnitude of
$r_{\mathrm AB} \sim 26$, which is $\sim 4$ magnitudes fainter
than the median magnitude of galaxies targeted with AAOmega.


\section{Data Reduction}\label{Sec:DataReduction}

We used version 4.66 of the 2dF data reduction
pipeline\footnote{http://www.aao.gov.au/AAO/2df/aaomega/} to process
the data taken with AAOmega.  The  steps for processing
  data from the red and blue arms are similar although not identical.
For both arms we define the spectral trace of each fibre using the
fibre flat fields, then extract spectra from the science frames, arcs
and flat fields, wavelength calibrate the extracted spectra from both
the science data and the fibre flats, flat field the extracted spectra
using the spectra from the fibre flats, and subtract the sky using the
spectra extracted form the sky fibres.

For the blue arm we first subtract a dark frame from the data to
reduce the impact of the bad columns in the blue detector, we remove
scattered light and we do the wavelength calibration with the arc. For
the red arm, we remove residuals from night sky lines using a PCA model
for the residuals and we use night sky lines for the wavelength calibration.

The data from the different configurations are then coadded. Not all
configurations were identical. For example, some targets were observed
in one configuration but not another, and some targets were
 observed with different fibres in different configurations. The pipeline keeps track of the objects that each
fibre targets, so coadding the spectra on an object--by--object basis
is straightforward. At this stage, we
use our own scripts, coded in Python and IRAF\footnote{IRAF is
  distributed by the National Optical Astronomy Observatories which
  are operated by the Association of Universities for Research in
  Astronomy, Inc., under the cooperative agreement with the National
  Science Foundation}, to coadd the spectra, since given the long
exposure times and the large number of exposures we needed better
control of selecting which spectra went into the final sum than what
is currently possible with the pipeline. The final step is to splice the
blue and red halves of each spectrum.

The redshifts are measured with Jan 08 version of RUNZ
\citep{Drinkwater2010}.  The most commonly identified features in the
spectra are emission lines from the [OII]\,$\lambda\lambda$\,3726,3728
and the [OIII] \,$\lambda\lambda$\,4959,5007 doublets, emission lines
from H--alpha and H--beta, and absorption lines from higher order Balmer
transitions and Ca II H and K. In a few spectra, we can identify lines
such as [NeIII]\,$\lambda$\,3869, [NII]\,$\lambda$\,6549,
[NII]\,$\lambda$\,6583 and the [SII]\,$\lambda\lambda$\,6716,6730
doublet. Spectra are assigned a redshift, a redshift uncertainty and a
 quality flag, following the scheme outlined in \citet{Drinkwater2010} and
listed in Table~\ref{table:flagdescription}. The
flag provides additional information about each spectrum, such as the
level of   assessment that the redshift is correct.  The reliability
of these quality flags is quantified in \citep{Drinkwater2010}, albeit in the context 
of their sample of highly star-forming emission line galaxies. Redshifts that have the
flag set to 4 or 5 are considered to be secure, since the redshifts
are obtained from at least 2 clearly identified features. Objects with
the flag set to 3  have less secure redshifts. Usually, these spectra have one
clear emission line, which we assume to be [OII]. Only spectra with
flags 3 or higher are considered in this paper.  In the
  following sections, we label objects that have a quality flag of 4 of
  5 as objects with secure redshifts and label objects that have a quality
  flag of 3 as objects with probable redshifts.

\begin{table*}[h]
\begin{center}
\caption{Description of redshift flags}\label{table:flagdescription}
\vspace{0.1cm}
\begin{tabular}{ll}
\hline\hline
Redshift Flag & Description \\
\hline                           
- & Not targeted with AAOmega\\
0 & Flawed spectrum \\
1 & Poor spectrum \\
2 & Possible redshift \\
3 & Probable redshift \\
4 & Secure redshift \\
5 & Secure redshift. High quality spectrum, more than 3 features clearly visible \\
6 & Not extragalactic \\
\hline
\end{tabular}
\end{center}
\end{table*}

\section{Results}\label{Sec:Results}


\subsection{Redshift completeness}

The percentage of targets for which we obtained a redshift with AAOmega
(the completeness fraction) is generally highest in the D4 field.  The
amount of time observing the D4 field was almost a factor of two
longer than the time used to observe the D1 field, so the higher
percentage is a reflection of the longer integration time.

The completeness fraction behaves similarly  for the two fields. It is
highest  for  photometrically classified  CC  SNe  that  did not  have
spectroscopic redshifts  and lowest for supernovae that  had neither a
spectroscopic redshift nor a  photometric one. Hence, in what follows,
we  pool  the  results  of  the two  fields  together  when  analysing
completeness as a function of magnitude, redshift and target type.

\begin{figure}
  \centering
  \includegraphics[width=8cm]{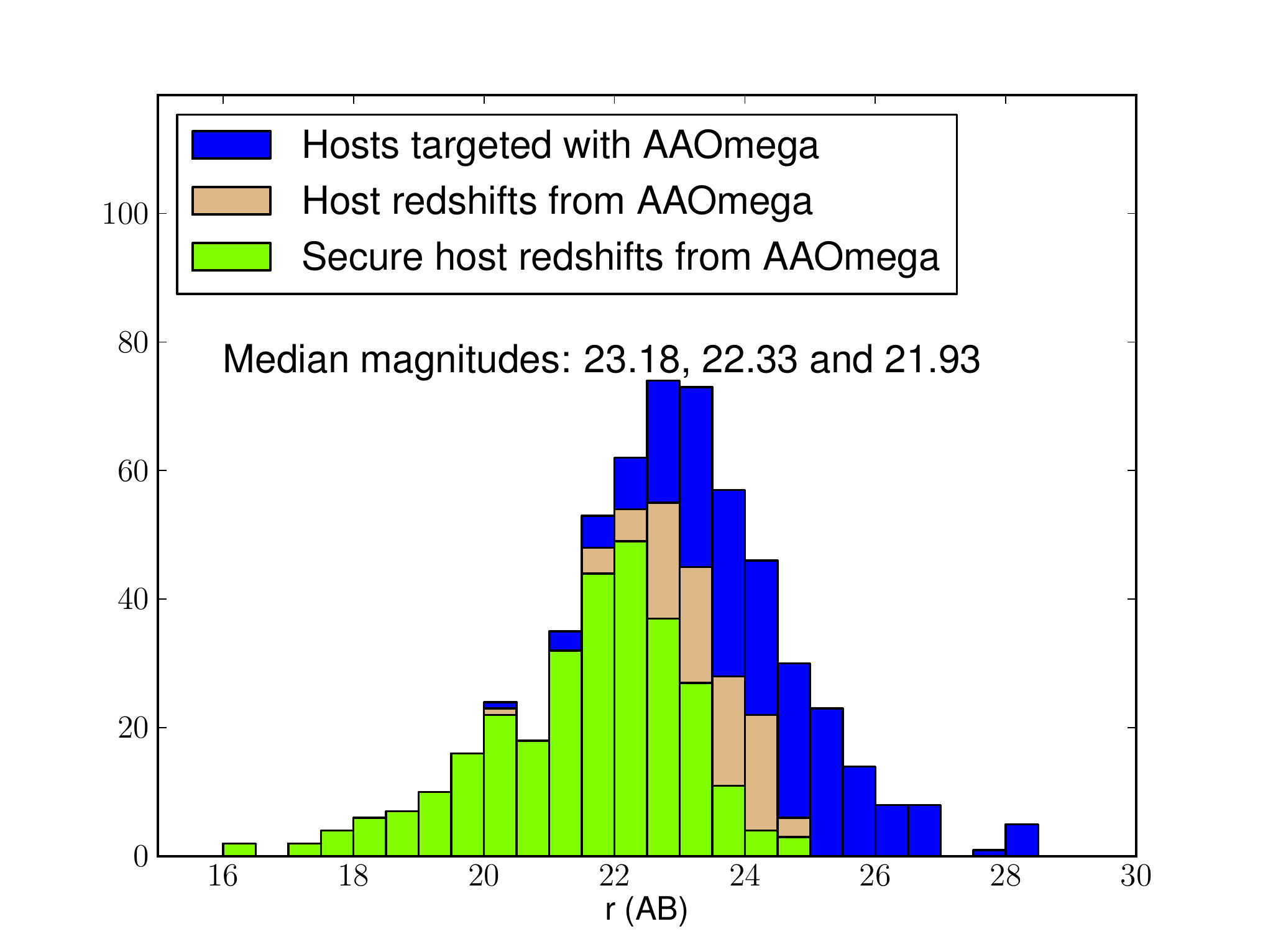}
 \includegraphics[width=8cm]{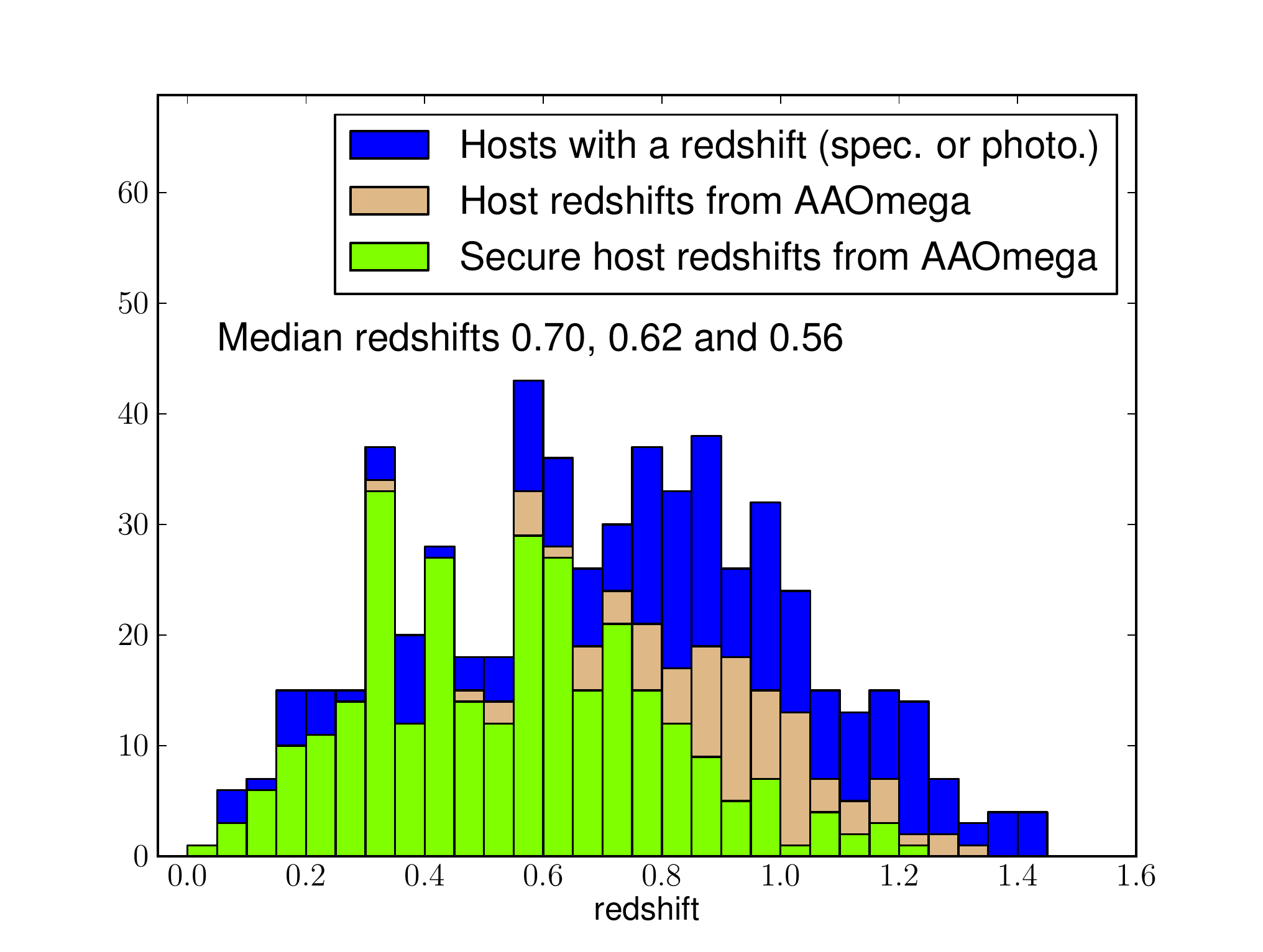}

  \caption{Histograms of the number of objects plotted as a function of
    magnitude {\bf (above)} and redshift {\bf (below)}. The
    blue, green and tan histograms represent all objects targeted with
    AAOmega, all objects with secure AAOmega redshifts   (quality flags 4 or
      5), and all objects with
    either a secure AAOmega redshift or a probable one  (quality flags 3, 4 or
      5). If the r-band
    magnitude of the host was unavailable, then the object was not
    plotted in the upper plot. If a spectroscopic redshift is
    not available for the lower plot, we use the photometric one. If
    neither a spectroscopic nor a photometric redshift were available,
    then the object was not plotted   in the lower panel. The median magnitudes
    and redshifts of objects in the blue, tan and green histograms
    annotate each figure.}

  \label{figure:Completeness}
\end{figure}

Histograms showing the number of galaxies with a redshift from AAOmega
are shown in Figure \ref{figure:Completeness}. Supernovae for which
the r--band magnitude of the host was unavailable are not shown.  Not
surprisingly, the median redshift of galaxies with a probable redshift
 (the quality flag set to 3) from AAOmega is higher than the
median redshift of galaxies that have a secure one. Usually, the [OII]
doublet is the only line that is detected in galaxies with probable
redshifts. Other lines, such as the [OIII] doublet, are not detected
because either they have been redshifted beyond the red end of the
spectral coverage (occurs by $z\sim 0.76$) or, less often, they are
too weak. At the resolution provided by AAOmega, the [OII] doublet is
only just resolved, which can sometimes help in making a redshift
marked as possible (quality flag 2) to one that is probable (quality
flag 3).

\begin{figure}
  \centering
  \includegraphics[width=8cm]{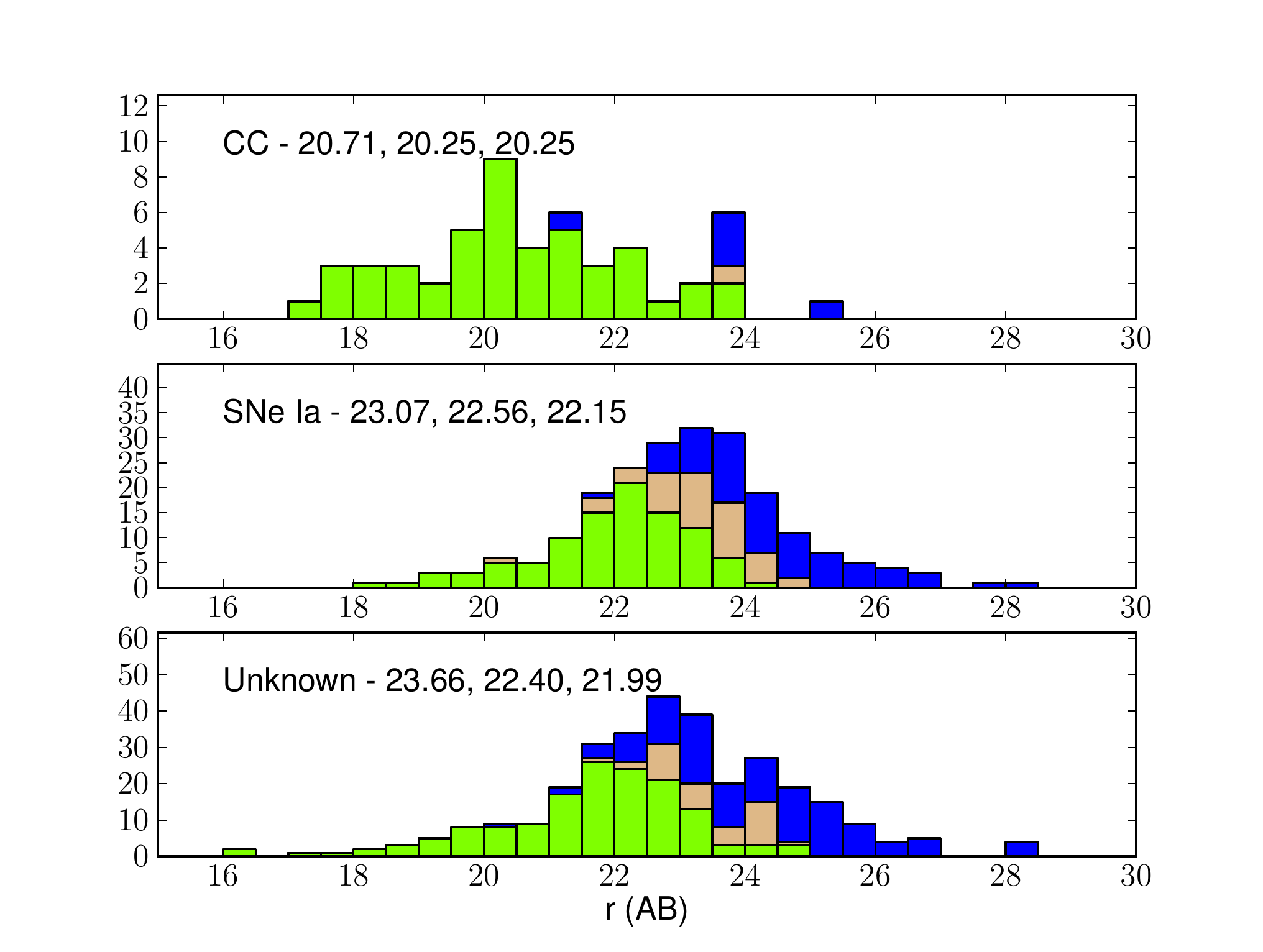}
  \includegraphics[width=8cm]{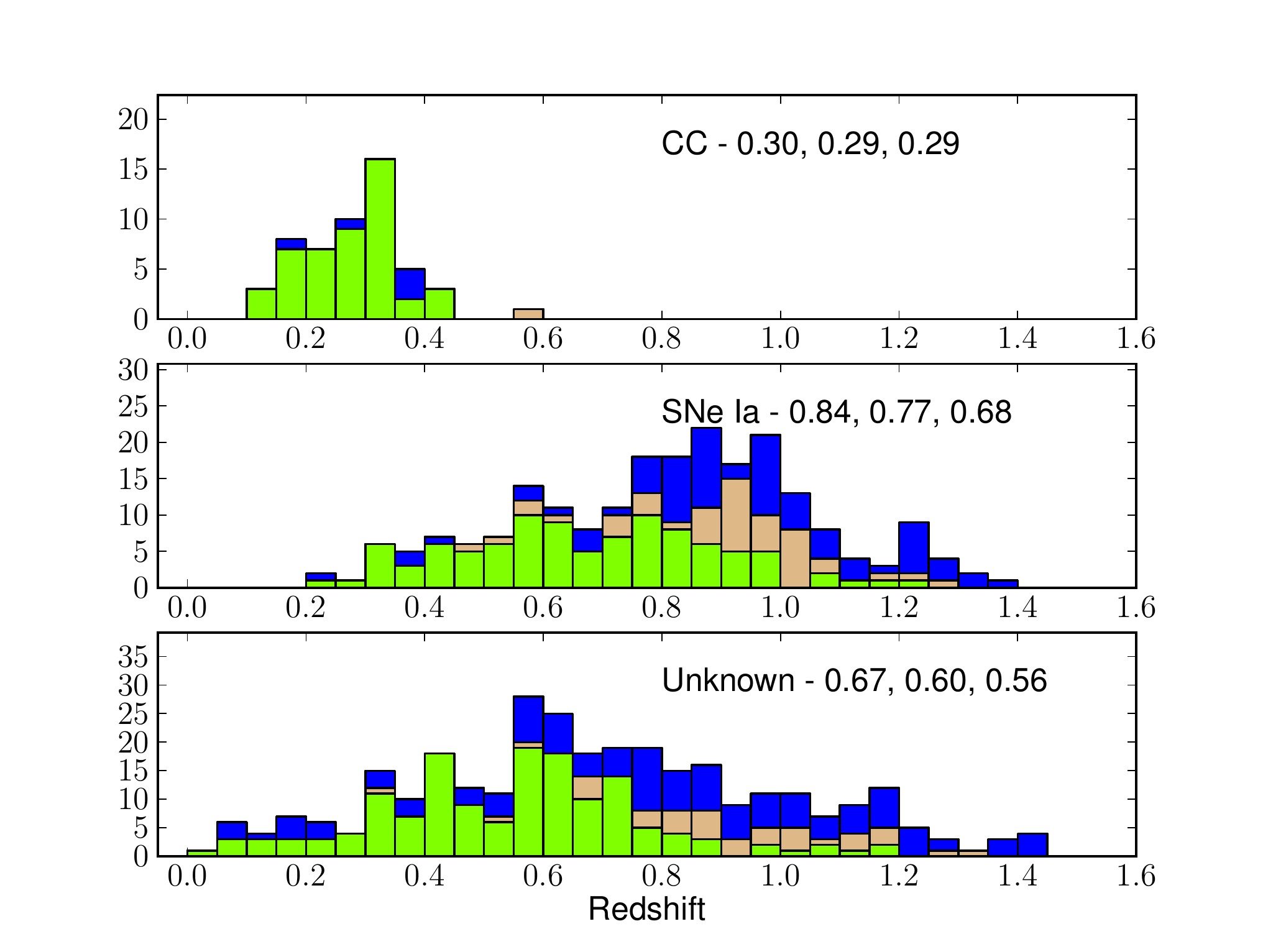}

  \caption{As for Figure \ref{figure:Completeness}, but split
    according to the photometric classification}

  \label{figure:Completeness2}
\end{figure}

In Figure~\ref{figure:Completeness2}, we split the magnitude and
redshift histograms according
to the photometric classification (i.e. SNe Ia, SNe CC or other
\citep[see][for details]{Bazin2011}). There are a couple of things to
note in these histograms.

In particular, the middle histogram in the lower plot shows that about
90\% of galaxies that hosted a SNe Ia at $z \sim 0.5$ have either a
secure redshift or a probable one. By $z \sim 1$, this
drops to about 50\%, of which about half are secure. Not shown in
this lower panel are supernovae that did not have either a spectroscopic
redshift (from any source) or a photometric one. These make up
about 15\% of all supernovae that were targeted with AAOmega.

The median redshift of objects photometrically classified as CC SNe is
lower than the median redshifts of the other two
classifications.  Primarily, this is driven by the
  absolute magnitude of CC SNe, which are, on average, about 2 mag
  fainter than SNe Ia.  The median apparent magnitude of galaxies
that hosted CC SNe is more that  one magnitude brighter,
leading to a relatively high completeness.

\subsection{White dwarfs}\label{Sec:WDObs}

In addition to allocating fibres to supernova hosts, we also allocated
fibres to candidate white dwarfs and to potential cluster galaxies. We
defer the discussion on the follow--up of cluster galaxies until the next
section: here we discuss the selection of white dwarfs.

White dwarfs can be easily identified in SNLS data, using either their
location in colour--colour plots or their relatively high proper motions. Both the
D1 and D4 fields were searched for white dwarfs using both techniques.
A listing of the candidates, notes on how they were found and the
spectral classification that we have assigned  to each of the objects are listed
in Table~\ref{table:whitedwarfs}. 

\citet{Limboz2008} also searched the SNLS fields for white dwarfs. The
criteria they use to select white dwarfs differ from ours, so their
catalogue and ours are not identical. They use different cuts in
colour--colour space, different magnitude limits and did not use proper
motions. However, nearly all of our candidates were also found by
them. The candidates in common are marked with the letter 'a' in
Table~\ref{table:whitedwarfs}. Since none of the \citet{Limboz2008}
candidates in the D1 and D4  fields had been spectroscopically confirmed, we
assigned fibres to all but one of objects that were in their
catalogue but not in ours. These objects are marked in
Table~\ref{table:whitedwarfs} with the letter 'b'. Not all white dwarf
candidates were targeted for spectroscopy. We avoided the brightest
one from \citet{Limboz2008} because of concerns that cross talk between
fibres would contaminate neighbouring spectra.

\begin{table*}[h]
\begin{center}
\caption{Candidate white dwarfs in the SNLS D1 and D4 fields}\label{table:whitedwarfs}
\vspace{0.1cm}
\begin{tabular}{rlcll}
\hline \hline 
RA & DEC & g band mag& Discovery method & Classification \\
\multicolumn{2}{c}{(J2000)} & (AB) &&\\
\hline
02 25 11.64    & -04 56 10.6$^a$ & 17.77 & Colours and proper motion & DA\\ 
02 24 05.27    & -04 22 16.3$^a$ & 17.90 & Proper motion only & DA \\ 
02 27 46.39    & -04 21 09.2$^a$ & 19.86 & Colours and proper motion & DC  \\
02 25 58.19    & -04 02 36.5$^a$ & 19.56 & Colours and proper motion & DA \\
&&&&\\
02 24 23.76    & -04 13 18.7$^b$ & 19.33 & Colours only & DA  \\
02 25 12.69    & -04 38 51.3$^b$ & 21.14 & Colours only & DA  \\
02 27 34.20    & -04 22 28.6$^b$ & 20.41 & Colours only & AGN $z=1.14$ \\
&&&&\\
22 14 58.39    & -18 07 42.6$^a$ & 20.39 & Colours and proper motion & DA \\ 
22 14 05.75    & -17 59 05.2$^a$ & 19.27 & Colours and proper motion & DAH? \\
22 16 21.93    & -17 44 07.9$^a$ & 20.31 & Colours only & DA\\
22 14 10.93    & -17 40 36.5$^a$ & 19.91 & Colours only & DA \\
22 13 26.42    & -17 23 18.9$^a$ & 20.37 & Colours and proper motion & DA \\
22 14 00.06    & -17 21 46.1     & 19.77 & Colours only & AGN $z=2.10$\\
&&&&\\
22 13 30.98    &-18 01 44.5$^b$ & 20.99  & Colours only & DC\\
\hline
\end{tabular}
\end{center}
\medskip
$^a$ Also listed in \citet{Limboz2008}\\
$^b$ From \citet{Limboz2008}\\
\end{table*}

Out of 14 candidates, 12 are white dwarfs and 2 are AGN. Of the 12
white dwarfs, we classify 9 as DA white dwarfs, two as DC white dwarfs
and one as a possible magnetic white dwarf (DAH). 
The two AGNs were both selected from their colours and were not in the
sample that used proper motions in the selection criteria. At high
galactic latitudes, AGN start to outnumber white dwarfs by $g=17$
\citep{Fan1999}, so they are an important source of
contamination. 

In Section~\ref{Sec:WD}, we discuss how DA white
dwarfs might be used as an alternative means of calibrating SN fluxes.

\subsection{Cluster galaxies}

For the D1 field, we assembled a catalog of 13 galaxies in 3 X--ray
selected clusters from the XMM--LSS survey \citep{Adami2011}, which
covers the entire D1 field. Out of 13 galaxies, 7 were assigned fibres
and 6 redshifts were obtained. The results are summarised in
Table~\ref{table:clusters}.  Both XLSS~J022404.1-041330 and
XLSS~J022303.3-043621 --- at $z=1.05$ and $z=1.22$, respectively --- had
been confirmed with data from other facilities \citep{Adami2011}. Our
observations reconfirm the redshift of XLSS~J022404.1-041330 but
failed to do so for XLSS~J022303.3-043621. Prior to our observations,
only one galaxy redshift had been measured for
XLSS~J022357.4-043516 \citep{Adami2011}. Our observations add three more redshifts to
this cluster.

\begin{table*}[h]
\begin{center}
\caption{Cluster redshifts}\label{table:clusters}
\vspace{0.1cm}
\begin{tabular}{lrrr}
  \hline   \hline
  Cluster                                & Cluster Class$^a$ & Mean redshift & Number of redshifts \\
  \hline
  XLSS~J022357.4-043516    & C1             & 0.495    & 4 \\
  XLSS~J022404.1-041330    & C1             & 1.051    & 1 \\
  XLSS~J022303.3-043621    & C2             & ...      & 0 \\
  \hline
\end{tabular}
\end{center}
\medskip
$^a$ See \citet{Pierre2007}\\
\end{table*}

\section{Discussion}\label{Sec:Discussion}

\subsection{A comparison with the real time follow--up}

It is instructive to compare the results we obtain here with the
results that we obtained with the VLT during the real time
follow--up of live candidates. During the final two years of the
spectroscopic follow--up of candidates with the VLT, the MOS modes of
FORS1 and FORS2 were used to observe both live candidates and the host
galaxies of other transients that were discovered in earlier years. By
the end of the SNLS survey, there were typically three to four host
galaxies visible in the FORS 7\arcmin\ by 7\arcmin\ FoV.

The MOS modes of FORS1 and FORS2 consists   of 38 movable blades that can
be used to make 19 slits anywhere in the FORS focal plane. While there
is a mode that allows pre--cut masks to be inserted into the 
FORS2 focal plane (called the MXU mode), this mode was not available for
Target--of--Opportunity requests.

 A number of targets were observed with both the VLT and
  AAOmega. Out of the 83 targets that were observed with both
  instruments, 63 have a redshift from both AAOmega and the VLT. The
  redshifts of all objects agree to within 0.002. The uncertainties in
  the redshifts from FORS1 and FORS2 are around 0.001 and are
  generally larger than the uncertainties in the redshifts from AAOmega.
  
\begin{figure}
  \centering
  \includegraphics[width=8cm]{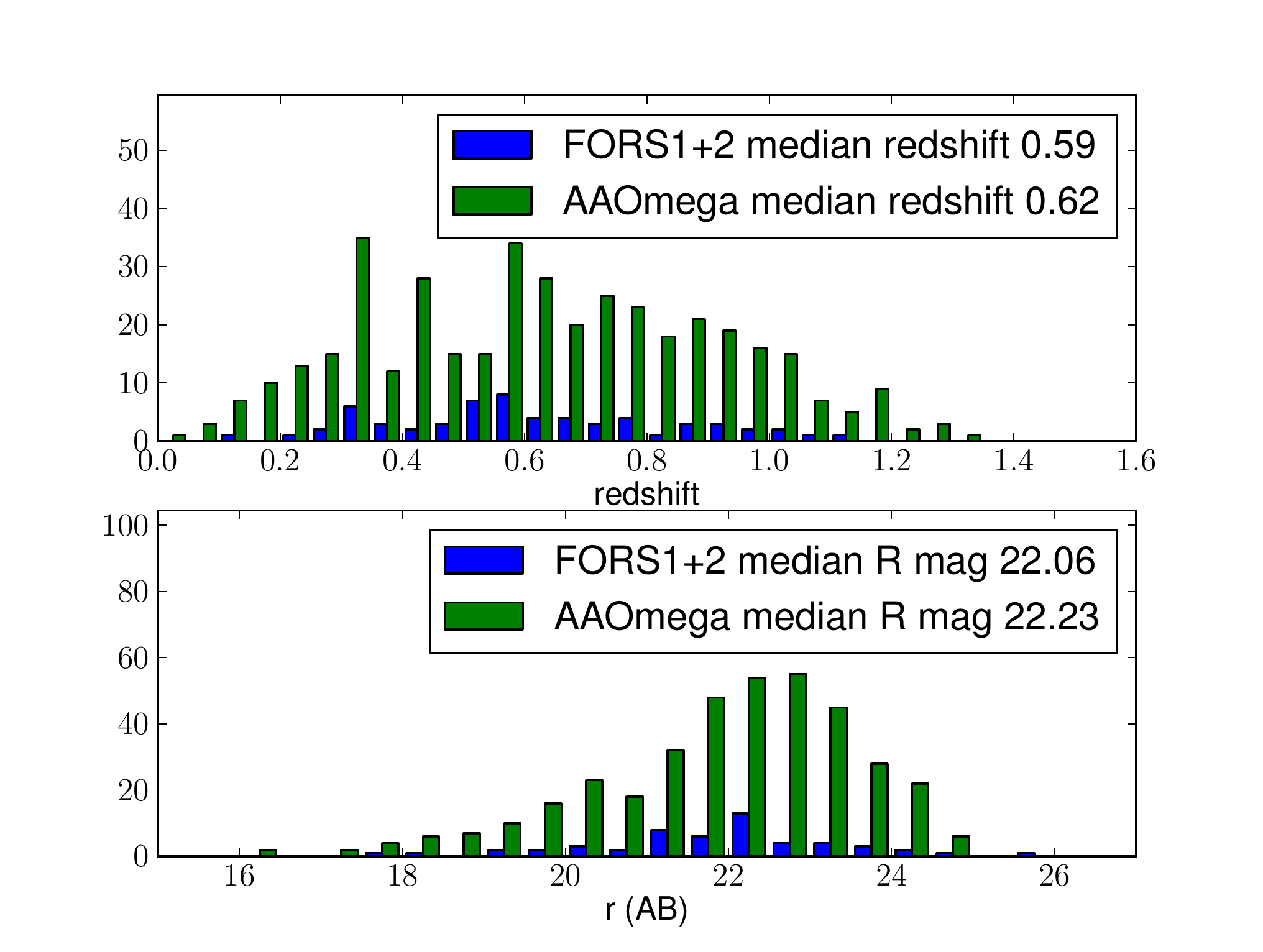}
  \caption{A comparison of the host redshift and magnitude distributions obtained with FORS1 and FORS2 with the host and magnitude distributions obtained with AAOmega.}
  \label{figure:FORSvsAAOmega}
\end{figure}

In Figure \ref{figure:FORSvsAAOmega}, we compare the magnitude and
redshift distributions of objects that were observed with FORS1 and
FORS2 with those that were observed with AAOmega. Excluding any sort of
renormalisation to account for differences in exposure times,
observing efficiency, and target selection, the redshift distributions are broadly similar.

The redshifts from FORS1 and FORS2 came from 66 separate MOS
setups. The exposure time for a single setup was typically one
hour. Summed over all setups, the total amount of time spent on the D1
and D4 fields with FORS1 and FORS2 was 240,000 seconds. The integration
times that were used with AAOmega for the D1 and D4 fields were 32,500
and 60,000 seconds, respectively.

At face value, it would seem that AAOmega --- a multi-object
spectrograph on a 4\,m class telescope --- has resulted in many more
redshifts than FORS1 and FORS2 --- both multi-object spectrographs on
8\,m class telescopes. While this is true, the difference is not as
extreme as that suggested by inspecting Figure
\ref{figure:FORSvsAAOmega}, since the number of targets available for
the AAOmega follow--up was about a factor of two larger than the
number available for FORS1 and FORS2. Nevertheless, it is difficult to
imagine that the efficiency with FORS could have been increased by
more than a factor of three if we were to repeat the experiment with
FORS2 with the MXU mode and with the catalogue that was used in the
follow--up with AAOmega. The main reason for the difference comes from
the difference in the FoV between AAOmega and FORS2. With AAOmega, we
can cover the entire 1 sq. degree MegaCam field in one shot, which is
60 times the area covered by FORS2. This more than compensates for
differences in telescope aperture and image quality.

While AAOmega is efficient in obtaining large numbers of host galaxy
redshifts, it is not suitable for confirming SN types --- which was,
together with obtaining redshifts, the main aim of the observations
	with the  VLT ---   over the redshift range that was
probed with FORS1 and FORS2.

\subsection{Redshifts: photometry vs. spectroscopy}

Photometric redshifts are often used as an
alternative to spectroscopic redshifts when samples become large
or when the precision and accuracy provided by photometric redshifts
are sufficient for the tasks at hand. They also allow redshifts to be
obtained for sources that are too faint for spectroscopy. For some
future transient surveys, such as the transient survey of the Large
Synoptic Survey Telescope (LSST), photometrically measured redshifts
of the hosts of the transients will be the only realistic option for
obtaining redshifts for most of the transients.

Photometric redshifts are both less precise and less accurate (in the
sense that they are biased) than spectroscopic redshifts. Up to $z=1$,
the precision of photometric redshifts, as measured by the normalised
median absolute deviation   -- $\langle|\Delta z|/
  (1+z)\rangle$, \citep{Ilbert2006} -- is typically
around 0.03. This can be compared to the level of uncertainty that is
typical for redshifts that are measured spectroscopically, which is
generally better than 0.001.

The lower precision appears amenable to solution by simply
increasing the number of SNe Ia in the sample. The issue of accuracy
is more critical, since any change in the accuracy with redshift might
be interpreted as evolution in the dark energy equation--of--state
parameter.

\begin{figure}
  \centering
  \includegraphics[width=8cm]{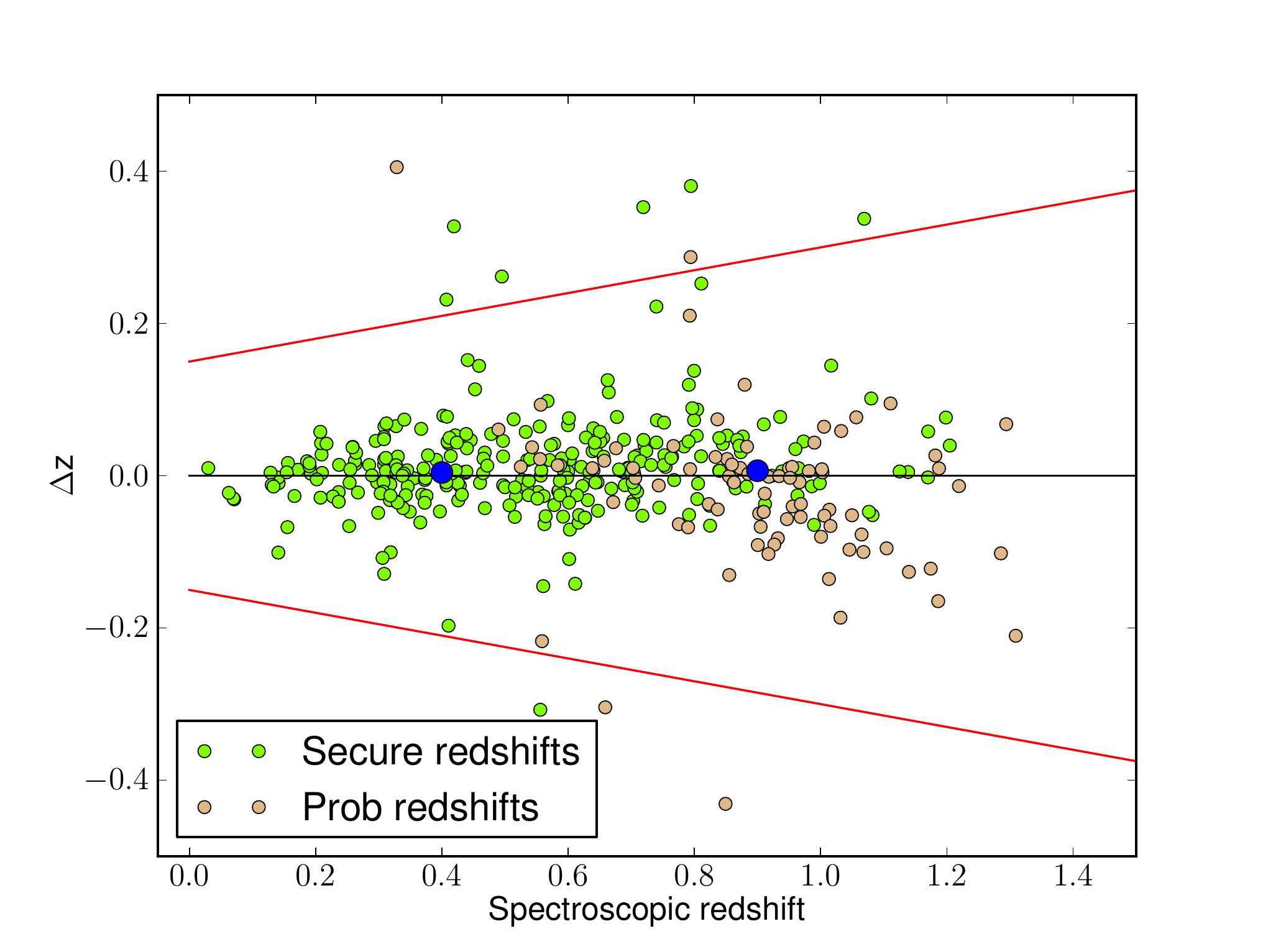}
  \caption{The deviation   $\Delta z$ between spectroscopic and photometric
    measures of redshift plotted against the spectroscopic redshift
    for objects with secure and probable spectroscopic redshifts (the
    green and tan circles respectively) . The median of the difference ---
    in bins of width 0.5 in redshift, centered at $z=0.4$ and
    $z=0.9$ --- are plotted as the large blue circles. The horizontal
    black line represents equality between the photometric and
    spectroscopic redshifts . The red lines represents the boundary
    beyond which the photometric redshift differs from the
    spectroscopic one by more than $0.15  (1+z)$.}
  \label{figure:photoZComp}
\end{figure}

In Figure \ref{figure:photoZComp}, we compare the photometric
redshifts (from \citet{Ilbert2006}) assigned to our targets with the spectroscopic
redshifts measured in this paper with AAOmega. Objects with secure and
probable redshifts are represented by the green and tan symbols,
respectively.  

To estimate the accuracy of the photometric redshifts, we compute the
median difference between spectroscopic and photometric measures of
the redshift in two broad redshift bins centered at $z=0.4$ and
$z=0.9$ and plot them as blue circles in
Figure~\ref{figure:photoZComp} . The values in these two bins are
0.004 and 0.007, respectively, with uncertainties, which are estimated
via bootstrap resampling, of 0.002 for both bins. In the standard
${\mathrm \Lambda}$CDM cosmology, an offset of 0.007 in redshift at
$z=0.9$ corresponds to a change of 1\% in luminosity, which
changes the best fit value for the dark energy equation--of--state
parameter by about 4\%. In SN Ia samples that are currently used for
cosmology, the systematic error in the dark energy equation--of--state
parameter is around 6\% \citep{Conley2011,Sullivan2011}. Samples that
use photometric redshifts for host galaxies instead of spectroscopic
redshifts will need to manage a source of systematic error that is
roughly similar in magnitude to the systematic error in current
samples.

In principle, one could remove the offset in the photometric redshifts
using the spectroscopic redshifts to work out the size of the offset,
as we have done here. With about 160 redshifts (the number of
redshifts that go into computing the location of the blue point at
$z=0.9$ in Figure \ref{figure:photoZComp}), the uncertainty in the
correction is 0.002, which at $z=0.9$ corresponds to a 1\% error in
the dark energy equation--of--state parameter. With significantly more
redshifts, the uncertainty could be made smaller still.

An interesting alternative to using redshifts of the host galaxies to
determine the size of the offset, which we do here, is to use
redshifts of general field galaxies, since these are more numerous.
However, there may be serious limitations to this approach. At some
level, the offset must depend on galaxy type. For example, the offset
for galaxies with strong emission lines will be different for galaxies
that lack emission lines. So estimates of the size of the offset may
be erroneous if the population of galaxies that hosted SNe Ia differs
from population of galaxies that is used to determine the offset.

 Photometric redshifts are also prone to catastrophic
  failures, which lead to non-Gaussian tails in the normalised
  redshift deviation: $\Delta z / (1+z)$. A commonly used definition for
  catastrophic failures is when the photometric redshift differs from
  the spectroscopic one by more than $0.15 (1+z)$
  \citep{Ilbert2006}. We can use our redshifts from AAOmega to make an
  independent measurement of the rate of catastrophic
  failures. Considering secure and probable redshifts together, the
  rate is 2.9\%, which is similar to the 3.7\% rate reported in
  \citet{Ilbert2006}.

\subsection{DA white dwarfs as in--situ flux calibrators}\label{Sec:WD}

Systematic errors now contribute as much to the uncertainty in the
dark energy equation--of--state parameter as statistical errors
\citep{Conley2011,Sullivan2011,Suzuki2011}. The largest systematic
error, by far, comes from calibration \citep{Conley2011,Sullivan2011}, which in
simple terms is the process of converting the observed counts in some
filter to a relative flux. For the purpose of constraining cosmology,
it is not necessary to know the absolute flux. 

DA (pure--hydrogen) white dwarfs are commonly used as flux
calibrators. For example, the primary spectrophotometric standards of
the {\it Hubble Space Telescope} (HST) are three DA white dwarfs
\citep{Bohlin2000}. The absolute precision of the spectral energy
distributions (SED) of these stars is around 1\% in the optical and
near--IR \citep{Bohlin2007}. The precision of DA white dwarf models,
which use spectral measurements of the effective gravity and
temperature, is thought to be even greater \citep{Allende--Prieto2009}. 

The idea of extending the HST network of primary standards to more DA
white dwarfs is explored in \citet{Allende--Prieto2009}. From an
initial sample of 598 objects from the Sloan Digital Sky Survey
(SDSS), they build a sample of 57 suitable DA white dwarfs (where
suitable means a sufficiently small
$\chi^2$ in the model fit to the SED, and model parameters $21 000 <
T_{\mathrm eff} < 85 000$ and $ 7.1 < \log g < 9$) that are then
analysed in more detail.  This sample is then reduced to 9 after
taking into account all sources of error, including extinction, which
becomes important for more distant objects.

In the two one--square--degree SNLS fields that we observed, we found 4
DA white dwarfs per field down to $g_{\mathrm AB} \sim 20$ (see
Section~\ref{Sec:WDObs}). For DES \citep{Bernstein2011}, which has a
field--of--view that is three times larger, there will be $\sim 12$ DA
white dwarfs to this magnitude limit. For SkyMapper
\citep{Keller2007}, which has an even larger field of view, there will
be around 25 DA white dwarfs in regions that have a similar galactic
latitude to the DES and SNLS fields, and considerably more at low
galactic latitudes.

While it is encouraging that there are a sufficient number of DA white
dwarfs for one to consider the possibility of using DA white dwarfs as
in--situ calibrators, the key issue is whether or not the space density
of suitable white dwarfs is sufficient. Scaling from the numbers
reported in \citet{Allende--Prieto2009} would suggest that the answer
to this is question is no. However, the needs for supernova cosmology
may be less stringent, since only the relative calibration of the SED
as a function of wavelength and the relative normalisation of the SEDs
matters. Furthermore one can average the results over the white dwarfs
that are within the field.

\subsection{Selection biases}\label{sec:biases}

The SNLS three--year constraints on the dark energy equation--of--state
parameter used SNe Ia that were discovered in real time and
subsequently confirmed to be SNe Ia with spectroscopy of the supernovae
near maximum light. For simplicity, we will refer to these SNe Ia as
the {\it SNLS three--year sample}. The--real time discovery of candidates and
the assignment of priorities for spectroscopic follow--up are described
in \citep{Howell2005, Sullivan2006a, Perrett2010}.

The criteria used to select and classify candidates in
\citet{Bazin2011} differs from those used for the SNLS three--year
sample . Furthermore, the spectroscopy undertaken with AAOmega is not
used to confirm the supernova type, as was done for SNe Ia in the SNLS
three--year sample, but to obtain the host redshift. Hence biases
incurred in using a SN Ia sample that use photometrically
classified   SNe Ia  with host galaxy redshifts (we
will refer to this sample as the {\it host--z sample}) to
constrain the dark energy equation--of--state will differ from the
biases that affect the SNLS three--year sample.

\citet{Perrett2010} make a detailed analysis of the biases incurred in
the SNLS three--year sample due to incomplete coverage of the
underlying SNe Ia population. In this section, our aim is not to
extend this work to the host--z sample but to describe how
the biases are likely to differ between the two samples. Nevertheless,
such a study will be necessary if the host--z sample is to
be used to constrain cosmological parameters.

Selection biases are incurred at three stages: the initial photometric
selection of candidates, the spectroscopic follow--up, and the
application of selection cuts to the fully assembled and integrated
data set. We briefly discuss the biases associated to the photometric
selection of candidates first, then discuss the biases associated to
the spectroscopic follow--up in more detail, and end with a comment on
the impact of selection cuts.

While the photometric selection of SNe Ia in the two samples differ,
the nature of the selection bias is the same for both
samples. Supernova are not perfect standard candles, hence magnitude
limited samples (as both these samples are) will tend to select the
brightest events as the magnitude limit of the survey is approached
(Malmquist bias). Once assumptions are made about the properties of
the parent SNe Ia population and how these properties evolve with
redshift, the bias can be computed with Monte--Carlo simulations
\citep{Perrett2010}. At a given redshift, the photometric
sample   of \citet{Bazin2011} is less biased than the SNLS three--year
sample, because it has a fainter magnitude limit.

The biggest difference between the two samples lies with the
spectroscopic follow--up.  For the SNLS three--year sample, the
spectroscopic follow--up can be broken into three steps: i) the
real--time sample is filtered for SNe Ia that are deemed to be
suitable for spectroscopy, ii) a request for spectroscopy is made and a
spectrum is taken, and iii) the supernova is identified.

The bias incurred from the last step is probably the most difficult to
model. One well known difficulty in positively identifying SNe Ia
spectroscopically is that SNe Ia in bright hosts are more difficult to
classify than SNe Ia in faint hosts. The difficulty increases with
redshift. The effect is similar to that imposed by the percentage
increase cut, which is used in the SNLS three--year sample. Imposing
this cut at the stage of selecting which candidates go on to be
observed spectroscopically makes computing the bias more tractable, as
part of the bias can be modelled using the imaging data.

The bias that results from the spectroscopic follow--up of host
galaxies in the host--z sample are different. It is generally a
lot easier to obtain the redshift of a galaxy if it is bright, if it
is forming stars or if it recently did so (such post--starburst
galaxies may lack strong emission lines but contain clear Balmer
absorption lines due to the presence of large numbers of A--type
stars). If the properties of SNe Ia were independent of the properties
of their hosts, then it would not matter if it was easier to get
redshifts for these galaxies.  However, there is ample evidence that
the properties of SNe Ia do depend on the properties of their
hosts. Brighter SNe Ia with broader light curves are preferentially
found in star forming galaxies \citep{Hamuy1995, Hamuy2000,
  Sullivan2006b}, and brighter SNe Ia also tend to live in more
massive (and therefore brighter) galaxies \citep[see][for
example]{Sullivan2010}.

This selection bias might be large enough to substantially affect the
accuracy of future experiments, such as the DES SN survey. However,
with a sufficiently large enough sample of host galaxies, it should be
possible to model the probability of getting a host redshift given the
photometric properties of the host.

There is also the possibility that the wrong host galaxy is
identified. Fortunately, this seems to be relatively
rare.  \citet{Dawson2009} show an example where the
  redshift of what appears to be the likely host disagrees with that
  of the supernova.  Another interesting example is SNLS-P04D1ay from
  this paper. SNLS-P04D1ay was observed with FORS2 and AAOmega. Two
  sets of lines at different redshifts ($z=0.446$ and $z=0.905$) could
  be identified in the spectra from both instruments. This is a clear case of two galaxies along the same
  line of sight. A potential way to eliminate some of these events
would be to exclude events in which the photometric redshift from
either the supernova or the host disagrees with the spectroscopic one
by an amount that does not exclude too many supernovae that have been
correctly associated to their hosts.

Finally, before the fully assembled SNe Ia data--set is used for
cosmology -- or for other studies, such as supernova rates -- selection
cuts  for quality are applied. Usually, this means that for every supernova there
needs to be a certain number of observations with a given
signal--to--noise ratio in a certain number of filters both before and
after maximum light. For the SNLS three--year sample, the selection cuts
reduce the number of SNe used for cosmology from 279 to 242
\citep{Conley2011}. Similar selection cuts 
  are applied to the host--z sample; however, these cuts usually
  occur during the initial selection.


\section{Summary and Conclusions}

Using the 2dF fibre positioner and the AAOmega spectrograph on the
AAT, we observed 698 galaxies in two SNLS fields that hosted
supernovae during the first three years of the SNLS. With integrations
varying from 10 to 60 ksec per object, we obtained redshifts for
\redshifts of them. The redshifts cover a very broad range, from
$z=0.03$ to $z=1.3$.

The percentage of fibres allocated to host galaxies was very high. Even
though the number of supernova hosts outnumbered the number of fibres
by as little as 10\%, we were able to allocate almost 100\% of the fibres
that were free to observe supernova hosts. This was
achieved with the additional restriction that all the hosts were
located in the central one square degree of the 2dF field--of--view.

The median redshift of photometrically classified SNe Ia in our sample
is \zm, which is higher than the median redshift of SNe Ia in the SNLS
3--year SNe Ia sample ($z=0.61$) and higher than the expected median
redshift of SNe Ia in the upcoming DES supernova survey.

The DES supernova search will discover around 4\,000 SNe Ia in the
redshift range $0.2 < z < 1.2$ over a period of approximately 5 years
\citep{Bernstein2011}.  The number of discoveries is an order of
magnitude larger than the number of SNe Ia found in either the SDSS
supernova survey or SNLS. Spread over 10 fields, this amounts to
 400 SNe Ia per field. After five years of operation, the
  number of SNe Ia per DES supernova field will be similar to the
  number of fibres in 2dF.

Obtaining the redshifts of galaxies that hosted photometrically
classified SNe with a multi--object fibre fed facility, such AAOmega on
the AAT, is an efficient alternative to obtaining a spectrum of every
supernova, one at a time, while it is bright enough to do so.
However, before photometrically classified SN Ia samples can be used
to constrain cosmology the biases in them need to be examined with the
level of rigour that is currently done for spectroscopically confirmed
samples.

\section*{Acknowledgments}

The authors thank the XMM--LSS collaboration, in particular Nicolas
Cleric and Marguerite Pierre, for providing the coordinates of
galaxies in three X--ray selected galaxy clusters in the D1 field.
Chris Lidman acknowledges the support from the Australian Research
Council (ARC) Future Fellowship Grant (FT0992259) and the excellent
support provided by the staff at the Australian Astronomical
Observatory. The Centre for All--sky Astrophysics (CAASTRO) is an
Australian Research Council Centre of Excellence, funded by grant
CE11E0090. Mark Sullivan acknowledges the support of the Royal
Society.

\section*{Appendix A1: Redshift Catalogue}

The redshift catalogue is presented in Table~\ref{table:redshifts}.
Only galaxies that have a redshift quality flag of 3 or higher are
listed. The supernovae that correspond to these galaxies are given
names that follow the convention {\it SNLS-PyyDnxx}. {\it SNLS-P} is
the prefix indicating that the supernova was found by
\citet{Bazin2011}, {\it yy} marks the year during which it occurred,
{\it Dn} marks the SNLS field in which it occurred (e.g.~D1 or D4),
and {\it xx} starts off as {\it aa} for the first supernova in a given
year and field and increments as {\it aa, ab, ..., az, ba, ...,} etc.~for new supernova. If the supernova matches an object in the real-time
database (RTD), the RTD name is also given. Redshift uncertainties are
generally less than 0.001. The coordinates refer to the location where
the fibre was placed. In most cases, this corresponds to the supernova
host. In cases where there was no clear host, the fibre was placed at
the location of the supernova.






\bibliographystyle{apj}
\bibliography{PASA_accepted}

\begin{table*}[h]
\begin{center}
\caption{AAOmega host galaxy redshifts of photometrically identified events}\label{table:redshifts}  
\vspace{0.1cm}
\begin{tabular}{llllll}
\hline \hline
Name   & RTD name & RA (J2000) & DEC (J2000) & Redshift & Quality flag \\
\hline
SNLS-P03D1ad & ... & 02 26 07.46 & -04 40 16.3 & 0.724 & 4\\
SNLS-P03D1af & ... & 02 24 28.52 & -04 08 38.2 & 1.083 & 4\\
SNLS-P03D1ah & SNLS-03D1be & 02 25 51.20 & -04 16 02.8 & 0.825 & 4\\
SNLS-P03D1ah & SNLS-03D1br & 02 26 09.42 & -04 23 45.9 & 0.426 & 5\\
SNLS-P03D1aj & SNLS-03D1am & 02 24 13.84 & -04 26 02.3 & 0.557 & 3\\
SNLS-P03D1ak & SNLS-03D1aq & 02 25 03.10 & -04 05 01.9 & 0.705 & 5\\
SNLS-P03D1al & SNLS-03D1bk & 02 26 27.49 & -04 32 11.7 & 0.866 & 4\\
SNLS-P03D1an & SNLS-03D1ax & 02 24 23.24 & -04 43 16.0 & 0.497 & 4\\
SNLS-P03D1ao & ... & 02 25 40.15 & -04 18 47.0 & 0.596 & 5\\
SNLS-P03D1ap & ... & 02 26 48.20 & -04 03 53.4 & 0.548 & 4\\
SNLS-P03D1ar & ... & 02 24 53.34 & -04 16 16.8 & 0.648 & 4\\
SNLS-P03D1ay & SNLS-03D1by & 02 27 54.02 & -04 03 04.4 & 0.378 & 5\\
SNLS-P03D1bg & ... & 02 24 47.80 & -04 31 02.0 & 0.584 & 3\\
SNLS-P03D1bh & SNLS-03D1df & 02 26 41.90 & -04 45 13.1 & 0.237 & 4\\
SNLS-P03D1bi & ... & 02 26 21.97 & -04 50 30.2 & 0.704 & 4\\
SNLS-P03D1br & ... & 02 24 42.89 & -04 21 03.2 & 1.294 & 3\\
SNLS-P03D1bu & ... & 02 24 24.26 & -04 44 26.4 & 0.142 & 5\\
SNLS-P03D1by & SNLS-03D1du & 02 24 59.84 & -04 18 38.2 & 0.852 & 3\\
SNLS-P03D1ca & ... & 02 26 51.58 & -04 14 15.0 & 0.210 & 4\\
SNLS-P03D1cb & ... & 02 25 21.28 & -04 23 00.0 & 0.602 & 4\\
SNLS-P03D1cd & SNLS-03D1fb & 02 27 12.84 & -04 07 16.7 & 0.497 & 4\\
SNLS-P03D1cg & ... & 02 27 48.49 & -04 19 01.6 & 0.689 & 4\\
SNLS-P03D1ck & SNLS-03D1ea & 02 27 50.35 & -04 05 02.2 & 0.313 & 5\\
SNLS-P03D1cp & ... & 02 27 06.13 & -04 31 28.3 & 0.986 & 4\\
SNLS-P03D1ct & ... & 02 24 29.11 & -04 09 55.0 & 0.265 & 5\\
SNLS-P03D1cu & ... & 02 24 32.61 & -04 30 03.2 & 0.328 & 5\\
SNLS-P03D1cv & ... & 02 25 08.57 & -04 38 25.1 & 0.366 & 5\\
SNLS-P03D1cw & SNLS-03D1gi & 02 25 18.12 & -04 31 55.7 & 0.525 & 3\\
SNLS-P03D4aa & ... & 22 16 37.85 & -17 55 20.5 & 0.562 & 5\\
SNLS-P03D4aa & SNLS-03D4at & 22 14 24.01 & -17 46 36.0 & 0.634 & 4\\
SNLS-P03D4ab & SNLS-03D4az & 22 15 47.79 & -18 07 51.2 & 0.409 & 5\\
SNLS-P03D4ab & SNLS-03D4ev & 22 16 51.39 & -17 20 03.1 & 0.538 & 4\\
SNLS-P03D4ac & ... & 22 13 54.79 & -17 17 11.4 & 0.408 & 5\\
SNLS-P03D4ac & ... & 22 16 54.51 & -18 02 12.1 & 0.936 & 4\\
SNLS-P03D4ad & SNLS-03D4da & 22 15 57.04 & -18 05 21.4 & 0.328 & 5\\
SNLS-P03D4ae & SNLS-03D4au & 22 16 09.92 & -18 04 39.2 & 0.469 & 5\\
SNLS-P03D4af & SNLS-03D4fd & 22 16 14.44 & -17 23 43.4 & 0.791 & 3\\
SNLS-P03D4ag & ... & 22 15 49.67 & -17 43 31.7 & 0.755 & 4\\
SNLS-P03D4ai & ... & 22 16 05.63 & -17 26 23.3 & 1.077 & 4\\
SNLS-P03D4aj & SNLS-03D4aa & 22 16 56.37 & -17 57 38.1 & 0.167 & 5\\
SNLS-P03D4ak & SNLS-03D4fy & 22 15 33.16 & -17 44 14.3 & 0.339 & 5\\
SNLS-P03D4al & ... & 22 15 26.85 & -17 16 00.3 & 0.643 & 5\\
SNLS-P03D4am & SNLS-03D4gg & 22 16 40.22 & -18 09 51.0 & 0.593 & 4\\
SNLS-P03D4an & SNLS-03D4bc & 22 15 28.21 & -17 49 48.1 & 0.573 & 4\\
SNLS-P03D4ao & SNLS-03D4ag & 22 14 45.85 & -17 44 21.9 & 0.285 & 5\\
SNLS-P03D4ao & SNLS-03D4gh & 22 15 34.59 & -17 56 10.4 & 0.341 & 5\\
SNLS-P03D4ap & SNLS-03D4hd & 22 15 39.43 & -17 53 37.4 & 0.872 & 4\\
SNLS-P03D4aq & SNLS-03D4ai & 22 16 41.54 & -17 26 10.4 & 0.202 & 5\\
SNLS-P03D4ar & SNLS-03D4be & 22 14 09.07 & -17 36 38.9 & 0.414 & 5\\
SNLS-P03D4as & ... & 22 14 41.49 & -17 32 13.7 & 0.338 & 4\\

\hline
\multicolumn{6}{r}{\it Continued on next page ...}
\end{tabular}
\end{center} 
\end{table*}

\pagebreak

\setcounter{table}{4}

\begin{table*}[h]
\begin{center}
\caption{\it --- continued from previous page}\label{table:redshifts}  
\vspace{0.1cm}
\begin{tabular}{llllll}
\hline \hline
Name   & RTD name & RA (J2000) & DEC (J2000) & Redshift & Quality flag \\
\hline
SNLS-P03D4au & ... & 22 15 43.69 & -17 48 35.6 & 0.800 & 4\\
SNLS-P03D4av & ... & 22 14 58.89 & -17 50 44.1 & 0.313 & 5\\
SNLS-P03D4aw & SNLS-03D4bj & 22 14 30.09 & -18 06 49.9 & 0.895 & 3\\
SNLS-P03D4ay & ... & 22 14 55.22 & -17 24 30.3 & 0.602 & 5\\
SNLS-P03D4az & SNLS-03D4bx & 22 14 48.64 & -17 31 17.7 & 0.600 & 4\\
SNLS-P03D4ba & SNLS-03D4bf & 22 17 01.34 & -18 00 42.9 & 0.805 & 5\\
SNLS-P03D4bb & SNLS-03D4cd & 22 17 18.42 & -18 01 26.4 & 0.845 & 4\\
SNLS-P03D4bc & SNLS-03D4bl & 22 14 05.44 & -17 18 38.1 & 0.319 & 5\\
SNLS-P03D4bf & SNLS-03D4cw & 22 16 11.46 & -17 13 59.7 & 0.155 & 5\\
SNLS-P03D4bh & SNLS-03D4cn & 22 16 34.56 & -17 16 13.6 & 0.818 & 3\\
SNLS-P03D4bi & ... & 22 16 38.23 & -17 36 50.8 & 1.006 & 3\\
SNLS-P03D4bl & ... & 22 17 10.80 & -17 24 08.0 & 1.205 & 4\\
SNLS-P03D4bt & SNLS-03D4cy & 22 13 40.47 & -17 40 55.1 & 0.927 & 3\\
SNLS-P03D4bu & SNLS-03D4cz & 22 16 41.80 & -17 55 34.5 & 0.697 & 4\\
SNLS-P03D4bw & SNLS-03D4ed & 22 16 19.76 & -17 31 27.5 & 0.860 & 3\\
SNLS-P03D4by & SNLS-03D4di & 22 14 10.24 & -17 30 24.3 & 0.905 & 3\\
SNLS-P03D4bz & SNLS-03D4dh & 22 17 31.08 & -17 37 47.6 & 0.627 & 5\\
SNLS-P03D4cb & ... & 22 13 31.21 & -17 30 14.2 & 1.310 & 3\\
SNLS-P03D4cd & SNLS-03D4ec & 22 14 43.70 & -17 21 40.6 & 1.016 & 3\\
SNLS-P03D4cf & ... & 22 13 27.50 & -17 40 05.0 & 0.368 & 4\\
SNLS-P03D4ch & SNLS-03D4dp & 22 15 39.04 & -17 16 52.4 & 0.955 & 3\\
SNLS-P03D4cj & ... & 22 15 16.12 & -17 20 48.2 & 0.540 & 5\\
SNLS-P03D4ck & SNLS-03D4gi & 22 17 08.68 & -18 12 46.4 & 0.578 & 4\\
SNLS-P03D4cl & ... & 22 16 58.24 & -17 18 32.8 & 1.068 & 3\\
SNLS-P04D1aa & SNLS-04D1ad & 02 24 15.96 & -04 28 44.9 & 0.392 & 5\\
SNLS-P04D1ab & ... & 02 27 08.48 & -04 11 16.9 & 0.157 & 4\\
SNLS-P04D1ac & ... & 02 25 09.52 & -04 10 41.5 & 1.003 & 3\\
SNLS-P04D1ac & SNLS-04D1ae & 02 25 57.33 & -04 33 21.8 & 0.617 & 4\\
SNLS-P04D1ag & SNLS-04D1ec & 02 24 15.25 & -04 23 08.6 & 0.603 & 4\\
SNLS-P04D1ai & SNLS-03D1gr & 02 25 59.00 & -04 09 28.2 & 0.296 & 5\\
SNLS-P04D1aj & ... & 02 27 01.41 & -04 15 30.7 & 0.528 & 5\\
SNLS-P04D1al & ... & 02 27 23.85 & -04 04 37.5 & 0.623 & 5\\
SNLS-P04D1am & ... & 02 25 37.74 & -04 35 49.1 & 0.612 & 4\\
SNLS-P04D1am & ... & 02 26 08.55 & -04 31 33.5 & 0.823 & 3\\
SNLS-P04D1an & SNLS-04D1fh & 02 26 59.38 & -04 29 42.3 & 0.687 & 4\\
SNLS-P04D1ap & SNLS-04D1de & 02 26 35.93 & -04 25 21.8 & 0.768 & 4\\
SNLS-P04D1au & SNLS-04D1hw & 02 25 13.39 & -04 52 02.0 & 0.350 & 4\\
SNLS-P04D1aw & ... & 02 27 39.16 & -04 11 02.1 & 0.718 & 4\\
SNLS-P04D1ax & SNLS-04D1hb & 02 26 20.26 & -04 07 45.5 & 0.912 & 3\\
SNLS-P04D1ay & SNLS-04D1dz & 02 27 53.68 & -04 05 12.1 & 0.446 & 5\\
SNLS-P04D1be & ... & 02 25 36.24 & -04 36 16.1 & 0.825 & 4\\
SNLS-P04D1bg & SNLS-04D1la & 02 26 39.80 & -04 49 53.0 & 0.319 & 4\\
SNLS-P04D1bi & SNLS-04D1hk & 02 24 45.94 & -04 24 05.1 & 0.635 & 4\\
SNLS-P04D1bj & SNLS-04D1iu & 02 24 31.30 & -04 03 29.6 & 0.254 & 5\\
SNLS-P04D1bk & SNLS-04D1jg & 02 26 12.64 & -04 08 07.1 & 0.585 & 4\\
SNLS-P04D1bl & SNLS-04D1jt & 02 25 47.62 & -04 22 32.8 & 0.639 & 5\\
SNLS-P04D1bn & SNLS-04D1oa & 02 25 11.76 & -04 10 49.4 & 1.006 & 3\\
SNLS-P04D1bo & SNLS-04D1or & 02 26 23.42 & -04 41 50.7 & 1.139 & 4\\
SNLS-P04D1bq & SNLS-04D1oe & 02 25 24.50 & -04 53 34.7 & 0.641 & 4\\
SNLS-P04D1br & SNLS-04D1ln & 02 25 53.45 & -04 27 02.9 & 0.208 & 5\\

\hline
\multicolumn{6}{r}{\it Continued on next page ...}
\end{tabular}
\end{center} 
\end{table*}

\pagebreak

\setcounter{table}{4}

\begin{table*}[h]
\begin{center}
\caption{\it --- continued from previous page}\label{table:redshifts}  
\vspace{0.1cm}
\begin{tabular}{llllll}
\hline \hline
Name   & RTD name & RA (J2000) & DEC (J2000) & Redshift & Quality flag \\
\hline

\hline
\multicolumn{6}{r}{\it Continued on next page ...}
\end{tabular}
\end{center} 
\end{table*}

\pagebreak

\setcounter{table}{4}

\begin{table*}[h]
\begin{center}
\caption{\it --- continued from previous page}\label{table:redshifts}  
\vspace{0.1cm}
\begin{tabular}{llllll}
\hline \hline
Name   & RTD name & RA (J2000) & DEC (J2000) & Redshift & Quality flag \\
\hline
SNLS-P04D1bs & SNLS-04D1kw & 02 25 09.96 & -04 15 13.9 & 0.901 & 3\\
SNLS-P04D1bv & SNLS-04D1qw & 02 26 25.60 & -04 37 24.6 & 0.312 & 5\\
SNLS-P04D1bw & SNLS-04D1ot & 02 26 22.42 & -04 31 13.1 & 1.014 & 3\\
SNLS-P04D1bx & SNLS-04D1nz & 02 25 15.65 & -04 41 01.3 & 0.264 & 5\\
SNLS-P04D1by & SNLS-04D1hi & 02 26 21.60 & -04 16 52.7 & 0.556 & 4\\
SNLS-P04D1ce & ... & 02 25 44.81 & -04 24 05.4 & 0.910 & 3\\
SNLS-P04D1ch & SNLS-04D1qa & 02 25 14.88 & -04 49 22.9 & 0.173 & 5\\
SNLS-P04D1cj & SNLS-04D1po & 02 25 24.55 & -04 21 09.0 & 0.141 & 5\\
SNLS-P04D1cm & SNLS-04D1oq & 02 24 39.56 & -04 15 56.6 & 0.330 & 5\\
SNLS-P04D1cq & ... & 02 26 27.24 & -04 33 53.8 & 0.795 & 4\\
SNLS-P04D1ct & SNLS-04D1pq & 02 24 52.84 & -04 07 40.8 & 1.017 & 4\\
SNLS-P04D1cu & ... & 02 25 51.16 & -04 59 17.6 & 1.001 & 3\\
SNLS-P04D1cy & SNLS-04D1pv & 02 24 32.48 & -04 24 19.9 & 0.873 & 3\\
SNLS-P04D1cz & SNLS-04D1qs & 02 25 44.81 & -04 24 05.4 & 0.910 & 3\\
SNLS-P04D1dc & SNLS-04D1qu & 02 26 18.65 & -04 04 38.1 & 0.343 & 4\\
SNLS-P04D1de & SNLS-04D1qn & 02 27 28.18 & -04 20 37.1 & 0.490 & 3\\
SNLS-P04D1df & SNLS-04D1pj & 02 26 50.27 & -04 12 08.2 & 0.156 & 5\\
SNLS-P04D1dg & SNLS-04D1pu & 02 27 28.43 & -04 44 41.5 & 0.639 & 3\\
SNLS-P04D1dk & SNLS-04D1pf & 02 25 13.69 & -04 57 49.2 & 0.311 & 5\\
SNLS-P04D1dm & ... & 02 24 10.34 & -04 38 15.3 & 0.647 & 4\\
SNLS-P04D1dn & ... & 02 24 20.81 & -04 30 35.7 & 0.640 & 4\\
SNLS-P04D1dq & SNLS-04D1qr & 02 25 49.10 & -04 29 00.2 & 0.429 & 4\\
SNLS-P04D1dw & SNLS-04D1sw & 02 26 44.52 & -04 16 39.0 & 0.710 & 4\\
SNLS-P04D1dx & SNLS-04D1tb & 02 24 09.59 & -04 50 11.8 & 0.495 & 4\\
SNLS-P04D1dz & SNLS-04D1rg & 02 26 43.64 & -04 33 17.8 & 0.309 & 5\\
SNLS-P04D1eg & SNLS-05D1ad & 02 24 54.23 & -04 43 47.6 & 0.408 & 5\\
SNLS-P04D1eh & SNLS-04D1sf & 02 25 56.14 & -04 30 44.7 & 0.707 & 3\\
SNLS-P04D1ej & SNLS-04D1sl & 02 24 20.96 & -04 03 50.1 & 0.703 & 3\\
SNLS-P04D1el & SNLS-04D1sc & 02 26 34.39 & -04 02 46.3 & 0.627 & 4\\
SNLS-P04D4aa & SNLS-04D4ih & 22 17 17.01 & -17 40 38.9 & 0.935 & 3\\
SNLS-P04D4aa & SNLS-04D4lj & 22 15 01.49 & -17 49 19.9 & 0.411 & 4\\
SNLS-P04D4ab & ... & 22 15 30.47 & -17 23 15.2 & 1.046 & 3\\
SNLS-P04D4ab & SNLS-04D4ay & 22 15 54.07 & -18 02 50.1 & 0.565 & 5\\
SNLS-P04D4ab & SNLS-04D4id & 22 16 21.50 & -17 13 44.2 & 0.770 & 4\\
SNLS-P04D4ac & ... & 22 17 26.04 & -18 01 44.9 & 0.715 & 5\\
SNLS-P04D4ac & SNLS-04D4kd & 22 13 44.03 & -17 57 58.9 & 0.745 & 5\\
SNLS-P04D4ad & SNLS-04D4lb & 22 14 59.53 & -18 01 46.1 & 0.800 & 4\\
SNLS-P04D4ae & SNLS-04D4mt & 22 16 54.10 & -17 31 56.0 & 0.371 & 5\\
SNLS-P04D4af & SNLS-04D4di & 22 15 40.78 & -17 40 12.7 & 0.947 & 3\\
SNLS-P04D4ag & SNLS-04D4it & 22 16 00.53 & -18 05 48.5 & 0.982 & 3\\
SNLS-P04D4ag & SNLS-04D4ms & 22 16 13.40 & -17 53 19.7 & 0.568 & 4\\
SNLS-P04D4ah & ... & 22 15 13.70 & -18 03 19.8 & 0.468 & 4\\
SNLS-P04D4ah & SNLS-04D4cm & 22 13 28.78 & -18 03 40.6 & 0.466 & 5\\
SNLS-P04D4ai & ... & 22 13 56.80 & -18 01 02.4 & 0.741 & 4\\
SNLS-P04D4ai & ... & 22 16 47.69 & -17 59 47.9 & 0.416 & 5\\
SNLS-P04D4aj & SNLS-04D4ev & 22 15 10.50 & -18 05 58.1 & 0.900 & 3\\
SNLS-P04D4ak & SNLS-04D4fu & 22 14 11.42 & -17 32 32.3 & 0.134 & 5\\
SNLS-P04D4al & SNLS-04D4dw & 22 16 44.63 & -17 50 02.4 & 1.032 & 3\\
SNLS-P04D4ao & SNLS-04D4im & 22 15 00.83 & -17 23 45.9 & 0.752 & 4\\
SNLS-P04D4ap & SNLS-04D4ec & 22 16 29.30 & -18 11 04.4 & 0.593 & 5\\

\hline
\multicolumn{6}{r}{\it Continued on next page ...}
\end{tabular}
\end{center} 
\end{table*}

\pagebreak

\setcounter{table}{4}

\begin{table*}[h]
\begin{center}
\caption{\it --- continued from previous page}\label{table:redshifts}  
\vspace{0.1cm}
\begin{tabular}{llllll}
\hline \hline
Name   & RTD name & RA (J2000) & DEC (J2000) & Redshift & Quality flag \\
\hline
SNLS-P04D4ap & SNLS-04D4in & 22 15 08.58 & -17 15 40.2 & 0.516 & 4\\
SNLS-P04D4aq & ... & 22 15 09.36 & -18 11 04.5 & 0.886 & 3\\
SNLS-P04D4aq & SNLS-04D4jw & 22 17 18.88 & -17 39 56.0 & 0.961 & 4\\
SNLS-P04D4ar & ... & 22 16 27.32 & -18 12 18.9 & 0.453 & 5\\
SNLS-P04D4as & SNLS-04D4eq & 22 14 23.63 & -17 52 04.6 & 0.701 & 5\\
SNLS-P04D4as & SNLS-04D4jv & 22 15 36.94 & -18 09 24.3 & 0.230 & 5\\
SNLS-P04D4at & ... & 22 14 44.87 & -18 09 20.0 & 1.171 & 4\\
SNLS-P04D4at & SNLS-04D4kq & 22 13 31.67 & -17 53 47.2 & 0.744 & 3\\
SNLS-P04D4au & SNLS-04D4kb & 22 17 16.25 & -17 38 37.8 & 0.967 & 3\\
SNLS-P04D4av & SNLS-04D4jx & 22 16 52.10 & -17 38 06.9 & 0.373 & 5\\
SNLS-P04D4aw & SNLS-04D4lg & 22 15 08.10 & -17 39 41.2 & 0.912 & 4\\
SNLS-P04D4ax & ... & 22 14 18.85 & -17 32 32.9 & 1.188 & 3\\
SNLS-P04D4ax & SNLS-04D4dm & 22 15 25.47 & -17 14 42.9 & 0.812 & 4\\
SNLS-P04D4ba & ... & 22 16 09.14 & -18 06 30.8 & 0.933 & 3\\
SNLS-P04D4bd & SNLS-04D4et & 22 14 51.77 & -17 47 23.0 & 0.536 & 5\\
SNLS-P04D4bd & SNLS-04D4kn & 22 15 04.42 & -17 19 45.6 & 0.910 & 4\\
SNLS-P04D4be & SNLS-04D4ew & 22 14 59.83 & -17 40 35.1 & 0.131 & 5\\
SNLS-P04D4be & SNLS-04D4jn & 22 14 27.43 & -17 58 12.6 & 0.330 & 5\\
SNLS-P04D4bf & SNLS-04D4eu & 22 14 47.34 & -17 46 11.6 & 0.795 & 3\\
SNLS-P04D4bh & SNLS-04D4ju & 22 17 02.72 & -17 19 58.7 & 0.472 & 5\\
SNLS-P04D4bi & SNLS-04D4fo & 22 14 47.10 & -17 25 16.1 & 0.880 & 3\\
SNLS-P04D4bk & SNLS-04D4fv & 22 16 04.73 & -17 59 31.9 & 0.533 & 4\\
SNLS-P04D4bk & SNLS-04D4ky & 22 17 27.99 & -18 13 37.1 & 0.559 & 3\\
SNLS-P04D4bl & ... & 22 15 45.53 & -17 32 14.3 & 0.918 & 3\\
SNLS-P04D4bl & SNLS-04D4mx & 22 16 30.26 & -18 03 41.5 & 1.220 & 3\\
SNLS-P04D4bm & ... & 22 16 21.79 & -18 04 04.6 & 0.852 & 4\\
SNLS-P04D4bm & SNLS-04D4lw & 22 15 53.40 & -17 52 58.6 & 0.850 & 3\\
SNLS-P04D4bo & ... & 22 16 58.55 & -17 56 14.9 & 0.740 & 5\\
SNLS-P04D4bp & ... & 22 14 40.64 & -17 35 37.6 & 0.720 & 4\\
SNLS-P04D4br & SNLS-04D4gd & 22 15 09.38 & -18 13 35.1 & 0.193 & 4\\
SNLS-P04D4bt & SNLS-04D4hk & 22 16 34.91 & -17 49 42.1 & 0.587 & 4\\
SNLS-P04D4bv & SNLS-04D4ft & 22 14 31.08 & -17 40 19.0 & 0.268 & 5\\
SNLS-P04D4bx & SNLS-04D4hg & 22 16 41.90 & -17 56 31.5 & 0.518 & 5\\
SNLS-P04D4by & ... & 22 16 48.93 & -17 53 20.8 & 0.805 & 4\\
SNLS-P04D4bz & ... & 22 13 56.35 & -17 49 45.6 & 1.199 & 5\\
SNLS-P04D4ca & SNLS-04D4gz & 22 16 58.96 & -17 37 18.4 & 0.375 & 5\\
SNLS-P04D4cb & ... & 22 14 03.62 & -17 37 57.3 & 1.002 & 3\\
SNLS-P04D4cc & SNLS-04D4gg & 22 16 09.23 & -17 17 40.1 & 0.424 & 5\\
SNLS-P04D4cf & ... & 22 16 52.33 & -18 02 13.8 & 0.804 & 4\\
SNLS-P04D4ch & SNLS-04D4hu & 22 15 36.21 & -17 50 20.3 & 0.703 & 5\\
SNLS-P04D4cj & SNLS-04D4ht & 22 14 33.27 & -17 21 31.5 & 0.218 & 5\\
SNLS-P04D4ck & SNLS-04D4hx & 22 13 40.55 & -17 23 03.5 & 0.543 & 3\\
SNLS-P04D4cl & SNLS-04D4ib & 22 16 41.70 & -18 06 18.3 & 0.705 & 4\\
SNLS-P05D1ac & SNLS-04D1ss & 02 26 48.35 & -04 27 50.8 & 0.328 & 5\\
SNLS-P05D1ac & SNLS-05D1cq & 02 25 19.09 & -04 47 33.7 & 0.309 & 4\\
SNLS-P05D1ad & ... & 02 26 05.11 & -04 38 07.7 & 0.563 & 5\\
SNLS-P05D1ae & ... & 02 25 44.29 & -04 41 29.7 & 0.672 & 3\\
SNLS-P05D1ae & SNLS-05D1cb & 02 26 57.12 & -04 07 03.1 & 0.632 & 4\\
SNLS-P05D1af & ... & 02 27 05.06 & -03 59 37.8 & 1.037 & 4\\
SNLS-P05D1ai & SNLS-05D1cf & 02 27 55.52 & -04 54 20.3 & 0.497 & 4\\

\hline
\multicolumn{6}{r}{\it Continued on next page ...}
\end{tabular}
\end{center} 
\end{table*}

\pagebreak

\setcounter{table}{4}

\begin{table*}[h]
\begin{center}
\caption{\it --- continued from previous page}\label{table:redshifts}  
\vspace{0.1cm}
\begin{tabular}{llllll}
\hline \hline
Name   & RTD name & RA (J2000) & DEC (J2000) & Redshift & Quality flag \\
\hline
SNLS-P05D1al & ... & 02 26 09.43 & -04 23 10.3 & 0.855 & 3\\
SNLS-P05D1aq & ... & 02 26 31.05 & -04 47 02.0 & 0.438 & 4\\
SNLS-P05D1as & SNLS-05D1ds & 02 27 38.14 & -04 39 33.8 & 0.309 & 5\\
SNLS-P05D1au & SNLS-05D1da & 02 26 24.54 & -04 26 15.2 & 0.426 & 5\\
SNLS-P05D1av & SNLS-05D1cs & 02 27 16.12 & -04 11 00.1 & 0.918 & 3\\
SNLS-P05D1ax & ... & 02 25 09.49 & -04 02 20.1 & 0.615 & 5\\
SNLS-P05D1ba & SNLS-05D1cm & 02 27 13.04 & -04 27 18.5 & 0.228 & 5\\
SNLS-P05D1bc & ... & 02 26 06.73 & -04 00 21.9 & 0.960 & 4\\
SNLS-P05D1bd & SNLS-05D1di & 02 25 08.46 & -04 14 06.7 & 0.629 & 4\\
SNLS-P05D1be & ... & 02 25 38.24 & -04 18 06.3 & 0.329 & 3\\
SNLS-P05D1bg & ... & 02 24 58.03 & -04 19 27.0 & 0.837 & 3\\
SNLS-P05D1bl & ... & 02 26 35.38 & -04 00 44.0 & 0.211 & 4\\
SNLS-P05D1bn & ... & 02 27 18.54 & -04 51 46.8 & 0.436 & 5\\
SNLS-P05D1bo & SNLS-05D1dd & 02 26 43.99 & -04 58 12.8 & 0.331 & 4\\
SNLS-P05D1bp & ... & 02 27 04.62 & -04 26 48.7 & 0.070 & 4\\
SNLS-P05D1br & ... & 02 26 28.91 & -04 07 13.9 & 0.884 & 3\\
SNLS-P05D1bv & SNLS-05D1je & 02 25 10.71 & -04 50 32.8 & 0.310 & 4\\
SNLS-P05D1bx & ... & 02 26 00.64 & -04 18 52.4 & 0.705 & 4\\
SNLS-P05D1bz & ... & 02 24 42.17 & -04 13 04.3 & 0.834 & 3\\
SNLS-P05D1ca & SNLS-05D1fa & 02 26 04.81 & -04 14 36.9 & 0.794 & 3\\
SNLS-P05D1cd & SNLS-05D1ej & 02 26 06.32 & -04 43 46.0 & 0.312 & 4\\
SNLS-P05D1cg & ... & 02 26 35.48 & -04 26 44.0 & 0.690 & 4\\
SNLS-P05D1cj & ... & 02 25 46.98 & -04 11 34.2 & 0.659 & 3\\
SNLS-P05D1co & ... & 02 26 59.77 & -04 34 30.3 & 1.175 & 3\\
SNLS-P05D1cp & SNLS-05D1fs & 02 27 29.33 & -04 22 48.9 & 0.346 & 4\\
SNLS-P05D1cq & ... & 02 26 33.35 & -04 00 38.7 & 0.345 & 4\\
SNLS-P05D1cu & ... & 02 24 15.86 & -04 35 20.5 & 1.182 & 3\\
SNLS-P05D1cw & ... & 02 27 34.29 & -04 38 11.4 & 0.585 & 4\\
SNLS-P05D1cx & ... & 02 27 05.30 & -04 37 46.0 & 0.071 & 4\\
SNLS-P05D1cz & ... & 02 27 07.55 & -04 23 58.5 & 0.237 & 5\\
SNLS-P05D1da & SNLS-05D1jv & 02 26 28.34 & -04 20 14.4 & 0.422 & 4\\
SNLS-P05D1dd & ... & 02 24 32.80 & -04 13 32.8 & 0.259 & 4\\
SNLS-P05D1de & ... & 02 27 55.92 & -04 54 57.4 & 1.140 & 3\\
SNLS-P05D1dg & ... & 02 25 46.23 & -04 31 29.4 & 1.057 & 3\\
SNLS-P05D1dh & ... & 02 26 28.26 & -04 00 51.7 & 0.209 & 5\\
SNLS-P05D1di & SNLS-05D1ib & 02 25 34.35 & -04 58 36.7 & 0.238 & 5\\
SNLS-P05D1dj & ... & 02 27 06.66 & -04 36 08.7 & 0.228 & 5\\
SNLS-P05D1dk & ... & 02 25 26.65 & -04 18 44.0 & 0.644 & 4\\
SNLS-P05D1dl & ... & 02 27 46.67 & -04 53 56.1 & 0.294 & 4\\
SNLS-P05D1dm & ... & 02 27 12.69 & -04 57 27.9 & 0.191 & 4\\
SNLS-P05D1dn & ... & 02 24 59.66 & -04 47 56.8 & 0.439 & 4\\
SNLS-P05D1do & SNLS-05D1hm & 02 27 46.16 & -04 43 02.1 & 0.586 & 4\\
SNLS-P05D1dp & SNLS-05D1ie & 02 24 23.58 & -04 40 39.4 & 0.263 & 5\\
SNLS-P05D1dr & ... & 02 26 41.52 & -04 00 01.8 & 0.877 & 4\\
SNLS-P05D1ds & ... & 02 25 23.57 & -04 05 10.4 & 0.600 & 4\\
SNLS-P05D1dt & ... & 02 27 29.96 & -04 49 31.9 & 0.695 & 4\\
SNLS-P05D1du & ... & 02 26 57.65 & -04 54 38.6 & 0.141 & 5\\
SNLS-P05D1dw & ... & 02 27 02.23 & -04 49 06.4 & 0.621 & 4\\
SNLS-P05D1ec & ... & 02 27 28.06 & -04 38 48.3 & 0.298 & 4\\
SNLS-P05D1ed & ... & 02 27 16.08 & -04 35 03.7 & 0.357 & 5\\

\hline
\multicolumn{6}{r}{\it Continued on next page ...}
\end{tabular}
\end{center} 
\end{table*}

\pagebreak

\setcounter{table}{4}

\begin{table*}[h]
\begin{center}
\caption{\it --- continued from previous page}\label{table:redshifts}  
\vspace{0.1cm}
\begin{tabular}{llllll}
\hline \hline
Name   & RTD name & RA (J2000) & DEC (J2000) & Redshift & Quality flag \\
\hline
SNLS-P05D1ee & ... & 02 27 02.49 & -04 35 31.2 & 0.030 & 5\\
SNLS-P05D1ef & SNLS-05D1jq & 02 24 12.10 & -04 41 41.8 & 0.500 & 5\\
SNLS-P05D1ei & ... & 02 27 16.41 & -04 22 52.3 & 0.514 & 4\\
SNLS-P05D1em & ... & 02 26 53.65 & -04 04 14.1 & 0.683 & 4\\
SNLS-P05D1en & ... & 02 25 39.89 & -04 05 25.0 & 0.063 & 5\\
SNLS-P05D1eq & SNLS-05D1jh & 02 24 25.48 & -04 03 47.4 & 0.765 & 4\\
SNLS-P05D1ev & SNLS-05D1ku & 02 26 13.23 & -04 24 14.7 & 0.319 & 5\\
SNLS-P05D1ey & SNLS-05D1kh & 02 25 56.06 & -04 13 29.4 & 0.327 & 5\\
SNLS-P05D1ez & ... & 02 26 03.81 & -04 42 09.8 & 0.603 & 4\\
SNLS-P05D1ff & ... & 02 27 46.42 & -04 08 25.9 & 0.874 & 4\\
SNLS-P05D1fi & ... & 02 24 55.14 & -04 53 35.2 & 0.710 & 4\\
SNLS-P05D1fj & ... & 02 26 26.00 & -04 54 46.7 & 0.658 & 3\\
SNLS-P05D1fm & ... & 02 25 45.14 & -04 55 23.0 & 0.903 & 3\\
SNLS-P05D1ft & ... & 02 24 00.31 & -04 04 33.2 & 0.571 & 4\\
SNLS-P05D1fv & ... & 02 27 55.05 & -04 49 56.9 & 0.676 & 3\\
SNLS-P05D1fw & ... & 02 27 11.32 & -04 50 20.4 & 0.964 & 4\\
SNLS-P05D1fx & ... & 02 27 15.90 & -04 49 46.3 & 0.722 & 4\\
SNLS-P05D1gk & ... & 02 24 17.30 & -04 55 03.6 & 0.965 & 4\\
SNLS-P05D1gm & ... & 02 27 30.80 & -04 40 13.9 & 0.578 & 5\\
SNLS-P05D1gn & ... & 02 27 18.37 & -04 33 47.9 & 0.309 & 5\\
SNLS-P05D1gu & ... & 02 25 00.92 & -04 17 29.2 & 0.863 & 4\\
SNLS-P05D1gv & ... & 02 24 58.70 & -04 17 38.7 & 0.254 & 4\\
SNLS-P05D1gw & ... & 02 24 46.34 & -04 18 29.5 & 0.669 & 4\\
SNLS-P05D1gy & ... & 02 24 05.31 & -04 19 25.4 & 0.557 & 4\\
SNLS-P05D1gz & ... & 02 24 03.91 & -04 18 04.5 & 0.466 & 5\\
SNLS-P05D1hb & ... & 02 27 32.99 & -04 00 30.0 & 0.706 & 4\\
SNLS-P05D1hc & ... & 02 27 27.07 & -04 07 20.5 & 0.255 & 5\\
SNLS-P05D1hd & ... & 02 27 04.24 & -04 05 17.3 & 0.767 & 3\\
SNLS-P05D1hf & ... & 02 25 22.38 & -04 09 31.9 & 0.558 & 5\\
SNLS-P05D1hg & ... & 02 24 09.98 & -04 13 16.6 & 1.050 & 3\\
SNLS-P05D1hi & SNLS-05D1ma & 02 24 19.58 & -04 08 04.0 & 0.432 & 4\\
SNLS-P05D1hk & SNLS-05D1lo & 02 25 54.37 & -04 20 04.7 & 0.862 & 4\\
SNLS-P05D4aa & SNLS-05D4ff & 22 16 20.16 & -18 02 33.0 & 0.403 & 5\\
SNLS-P05D4ab & SNLS-05D4ek & 22 16 27.49 & -17 44 09.3 & 0.536 & 4\\
SNLS-P05D4ac & SNLS-05D4al & 22 15 44.81 & -17 51 59.3 & 0.308 & 5\\
SNLS-P05D4ac & SNLS-05D4fl & 22 15 01.83 & -17 56 36.3 & 0.308 & 5\\
SNLS-P05D4ad & SNLS-05D4bc & 22 17 22.47 & -18 08 46.1 & 0.127 & 5\\
SNLS-P05D4ad & SNLS-05D4fg & 22 16 41.34 & -17 35 44.9 & 0.840 & 4\\
SNLS-P05D4ae & SNLS-05D4bf & 22 16 26.36 & -18 13 53.6 & 0.553 & 5\\
SNLS-P05D4af & ... & 22 16 56.54 & -17 52 11.5 & 0.745 & 5\\
SNLS-P05D4af & ... & 22 17 04.45 & -18 12 20.9 & 0.618 & 4\\
SNLS-P05D4ag & ... & 22 16 53.55 & -17 57 58.2 & 0.740 & 4\\
SNLS-P05D4ag & SNLS-05D4ay & 22 14 33.16 & -17 46 03.2 & 0.409 & 4\\
SNLS-P05D4ah & SNLS-05D4be & 22 16 53.39 & -17 14 10.1 & 0.538 & 4\\
SNLS-P05D4ai & SNLS-05D4bn & 22 16 07.41 & -17 30 40.7 & 0.878 & 4\\
SNLS-P05D4ai & SNLS-05D4fi & 22 16 09.58 & -17 14 16.2 & 0.411 & 5\\
SNLS-P05D4aj & ... & 22 14 40.11 & -17 37 35.2 & 0.657 & 4\\
SNLS-P05D4ak & SNLS-05D4av & 22 14 10.49 & -17 54 42.6 & 0.508 & 4\\
SNLS-P05D4al & ... & 22 14 07.08 & -17 39 21.8 & 0.793 & 3\\
SNLS-P05D4am & SNLS-05D4bl & 22 16 47.53 & -17 16 44.0 & 1.286 & 3\\

\hline
\multicolumn{6}{r}{\it Continued on next page ...}
\end{tabular}
\end{center} 
\end{table*}

\pagebreak

\setcounter{table}{4}

\begin{table*}[h]
\begin{center}
\caption{\it --- continued from previous page}\label{table:redshifts}  
\vspace{0.1cm}
\begin{tabular}{llllll}
\hline \hline
Name   & RTD name & RA (J2000) & DEC (J2000) & Redshift & Quality flag \\
\hline
SNLS-P05D4an & SNLS-05D4bi & 22 15 56.53 & -17 59 09.7 & 0.776 & 3\\
SNLS-P05D4ao & SNLS-05D4fp & 22 14 32.56 & -17 24 30.0 & 1.014 & 3\\
SNLS-P05D4ap & ... & 22 16 47.45 & -18 12 15.8 & 0.973 & 4\\
SNLS-P05D4aq & ... & 22 16 51.11 & -17 45 55.3 & 0.939 & 4\\
SNLS-P05D4ar & ... & 22 15 37.70 & -17 14 10.9 & 0.923 & 4\\
SNLS-P05D4ar & SNLS-05D4ar & 22 14 30.52 & -18 02 19.6 & 0.191 & 5\\
SNLS-P05D4as & ... & 22 14 35.03 & -17 25 15.9 & 0.479 & 5\\
SNLS-P05D4at & ... & 22 17 02.83 & -17 21 31.6 & 1.030 & 3\\
SNLS-P05D4at & SNLS-05D4jq & 22 14 50.86 & -17 33 23.3 & 1.081 & 4\\
SNLS-P05D4au & ... & 22 14 47.72 & -17 36 19.2 & 1.187 & 3\\
SNLS-P05D4aw & ... & 22 15 23.09 & -18 13 57.6 & 0.685 & 3\\
SNLS-P05D4ax & ... & 22 16 14.60 & -18 09 00.3 & 0.840 & 4\\
SNLS-P05D4ax & SNLS-05D4jp & 22 13 44.29 & -17 49 55.2 & 0.883 & 4\\
SNLS-P05D4ay & SNLS-05D4bw & 22 14 31.39 & -18 01 31.5 & 0.259 & 5\\
SNLS-P05D4ay & SNLS-05D4gx & 22 13 55.94 & -17 32 04.0 & 0.300 & 5\\
SNLS-P05D4az & ... & 22 13 28.98 & -17 52 49.1 & 0.792 & 4\\
SNLS-P05D4ba & SNLS-05D4bm & 22 17 04.60 & -17 40 39.9 & 0.373 & 5\\
SNLS-P05D4bb & ... & 22 14 33.00 & -17 17 38.2 & 0.420 & 5\\
SNLS-P05D4bc & SNLS-05D4ck & 22 16 01.52 & -18 04 59.2 & 0.459 & 5\\
SNLS-P05D4bd & SNLS-05D4ch & 22 14 13.84 & -17 22 42.4 & 0.856 & 3\\
SNLS-P05D4be & ... & 22 13 57.64 & -18 12 36.6 & 0.397 & 4\\
SNLS-P05D4be & SNLS-05D4jz & 22 16 05.43 & -17 42 16.0 & 0.516 & 4\\
SNLS-P05D4bg & SNLS-05D4cw & 22 14 50.05 & -17 44 19.9 & 0.376 & 4\\
SNLS-P05D4bg & SNLS-05D4gy & 22 16 28.50 & -17 51 49.4 & 0.588 & 4\\
SNLS-P05D4bh & SNLS-05D4jy & 22 16 22.93 & -17 46 59.4 & 0.868 & 4\\
SNLS-P05D4bi & SNLS-05D4de & 22 15 46.38 & -17 27 49.8 & 0.644 & 4\\
SNLS-P05D4bi & SNLS-05D4lb & 22 13 54.16 & -17 24 40.2 & 0.797 & 4\\
SNLS-P05D4bj & SNLS-05D4cb & 22 15 57.36 & -17 54 18.8 & 0.208 & 5\\
SNLS-P05D4bj & SNLS-05D4kp & 22 14 31.44 & -17 57 50.4 & 0.409 & 5\\
SNLS-P05D4bk & SNLS-05D4ca & 22 14 11.32 & -17 48 15.2 & 0.607 & 5\\
SNLS-P05D4bk & SNLS-05D4is & 22 17 08.06 & -18 01 30.6 & 0.837 & 3\\
SNLS-P05D4bl & SNLS-05D4cn & 22 13 31.43 & -17 17 20.8 & 0.764 & 4\\
SNLS-P05D4bl & SNLS-05D4jw & 22 17 14.21 & -17 55 08.5 & 0.309 & 4\\
SNLS-P05D4bm & SNLS-05D4cq & 22 14 09.68 & -18 13 34.6 & 0.701 & 4\\
SNLS-P05D4bm & SNLS-05D4jx & 22 14 55.55 & -17 30 23.2 & 0.811 & 4\\
SNLS-P05D4bn & SNLS-05D4dd & 22 16 17.23 & -17 22 31.5 & 0.791 & 4\\
SNLS-P05D4bo & ... & 22 13 57.28 & -18 02 47.7 & 0.590 & 4\\
SNLS-P05D4br & SNLS-05D4dt & 22 14 25.85 & -17 40 15.9 & 0.407 & 5\\
SNLS-P05D4bs & SNLS-05D4dr & 22 15 51.17 & -18 06 10.4 & 1.033 & 3\\
SNLS-P05D4bt & ... & 22 16 10.33 & -17 55 31.7 & 0.441 & 5\\
SNLS-P05D4bu & SNLS-05D4dp & 22 15 55.41 & -17 29 06.1 & 0.340 & 5\\
SNLS-P05D4bx & ... & 22 13 49.29 & -17 25 27.7 & 0.952 & 3\\
SNLS-P05D4by & SNLS-05D4dn & 22 16 01.37 & -17 57 17.3 & 0.191 & 5\\
SNLS-P05D4bz & SNLS-05D4dx & 22 13 39.37 & -18 03 21.0 & 0.792 & 4\\
SNLS-P05D4ca & SNLS-05D4eq & 22 14 07.74 & -17 52 35.8 & 0.950 & 3\\
SNLS-P05D4cb & SNLS-05D4en & 22 14 13.79 & -17 16 60.0 & 0.422 & 5\\
SNLS-P05D4cc & SNLS-05D4du & 22 15 29.58 & -17 54 05.4 & 0.306 & 5\\
SNLS-P05D4cd & SNLS-05D4fc & 22 16 53.04 & -17 41 13.4 & 1.065 & 3\\
SNLS-P05D4ce & ... & 22 15 07.24 & -17 16 58.5 & 0.969 & 3\\
SNLS-P05D4ch & ... & 22 15 37.65 & -18 04 21.7 & 1.171 & 4\\

\hline
\multicolumn{6}{r}{\it Continued on next page ...}
\end{tabular}
\end{center} 
\end{table*}

\pagebreak

\setcounter{table}{4}

\begin{table*}[h]
\begin{center}
\caption{\it --- continued from previous page}\label{table:redshifts}  
\vspace{0.1cm}
\begin{tabular}{llllll}
\hline \hline
Name   & RTD name & RA (J2000) & DEC (J2000) & Redshift & Quality flag \\
\hline
SNLS-P05D4ci & SNLS-05D4er & 22 16 04.12 & -17 34 00.3 & 1.069 & 4\\
SNLS-P05D4cl & SNLS-05D4ej & 22 15 52.51 & -18 11 44.3 & 0.586 & 5\\
SNLS-P06D1ac & SNLS-06D1ac & 02 24 07.83 & -04 55 52.7 & 0.416 & 4\\
SNLS-P06D1ag & ... & 02 24 21.65 & -04 40 08.5 & 0.641 & 5\\
SNLS-P06D1ah & SNLS-05D1lz & 02 25 27.53 & -04 28 42.8 & 0.461 & 5\\
SNLS-P06D1ai & ... & 02 25 14.10 & -04 16 59.8 & 0.863 & 3\\
SNLS-P06D1aj & SNLS-05D1ln & 02 24 45.17 & -04 20 37.2 & 0.656 & 4\\
SNLS-P06D1am & SNLS-05D1lj & 02 25 52.16 & -04 09 40.2 & 0.956 & 3\\
SNLS-P06D4aa & ... & 22 16 21.09 & -17 27 54.1 & 0.555 & 5\\
SNLS-P06D4ac & SNLS-06D4bw & 22 15 03.71 & -17 52 60.0 & 0.732 & 5\\
SNLS-P06D4ad & ... & 22 15 10.87 & -17 32 29.1 & 0.556 & 3\\
SNLS-P06D4af & ... & 22 13 37.14 & -17 35 45.5 & 0.991 & 3\\
SNLS-P06D4ag & ... & 22 16 56.62 & -17 25 34.0 & 1.267 & 3\\
SNLS-P06D4ah & SNLS-06D4bz & 22 13 44.15 & -17 17 20.5 & 0.663 & 4\\
SNLS-P06D4ai & SNLS-06D4cd & 22 16 39.20 & -18 00 03.8 & 0.590 & 5\\
SNLS-P06D4an & SNLS-06D4bo & 22 15 28.05 & -17 24 34.0 & 0.552 & 4\\
SNLS-P06D4ao & SNLS-06D4cb & 22 15 16.34 & -17 37 00.8 & 0.440 & 5\\
SNLS-P06D4ar & SNLS-06D4dp & 22 14 35.23 & -17 17 38.8 & 1.126 & 4\\
SNLS-P06D4as & ... & 22 17 34.79 & -17 56 55.1 & 0.840 & 4\\
SNLS-P06D4av & ... & 22 15 43.58 & -18 10 48.3 & 0.561 & 4\\
SNLS-P06D4aw & SNLS-06D4co & 22 15 26.49 & -17 52 09.0 & 0.439 & 4\\
SNLS-P06D4ax & SNLS-06D4dl & 22 16 59.80 & -17 15 02.9 & 0.753 & 4\\
SNLS-P06D4ay & ... & 22 16 24.10 & -17 50 44.7 & 1.161 & 3\\
SNLS-P06D4az & SNLS-06D4du & 22 15 47.37 & -17 54 41.9 & 0.602 & 4\\
SNLS-P06D4ba & ... & 22 15 55.56 & -17 41 36.3 & 0.665 & 4\\
SNLS-P06D4bb & SNLS-06D4cm & 22 15 15.75 & -17 18 11.8 & 0.918 & 3\\
SNLS-P06D4bc & SNLS-06D4ds & 22 16 01.40 & -18 05 20.7 & 0.318 & 5\\
SNLS-P06D4bd & SNLS-06D4cq & 22 16 55.48 & -17 42 43.4 & 0.413 & 5\\
SNLS-P06D4be & SNLS-06D4dt & 22 14 05.87 & -17 35 39.7 & 1.105 & 3\\
SNLS-P06D4bf & ... & 22 15 35.59 & -18 01 08.7 & 0.369 & 5\\
SNLS-P06D4bj & SNLS-06D4eb & 22 15 45.04 & -18 04 24.9 & 0.652 & 4\\
SNLS-P06D4bk & SNLS-06D4dv & 22 15 31.90 & -18 10 33.6 & 0.129 & 5\\
SNLS-P06D4bm & SNLS-06D4dh & 22 14 19.57 & -17 35 05.2 & 0.304 & 5\\
SNLS-P06D4bo & SNLS-06D4do & 22 15 48.60 & -17 17 00.7 & 0.290 & 5\\
SNLS-P06D4br & SNLS-06D4el & 22 16 27.25 & -17 51 22.7 & 0.587 & 4\\
SNLS-P06D4bs & ... & 22 14 59.91 & -17 31 37.1 & 1.166 & 3\\
SNLS-P06D4bt & ... & 22 16 35.03 & -17 25 45.7 & 0.784 & 4\\
SNLS-P06D4bv & ... & 22 14 33.12 & -17 19 54.8 & 0.752 & 4\\
SNLS-P06D4bw & SNLS-06D4ee & 22 13 29.35 & -17 55 47.6 & 0.651 & 4\\
SNLS-P06D4bx & ... & 22 14 10.66 & -18 08 44.2 & 1.111 & 3\\
SNLS-P06D4bz & SNLS-06D4gq & 22 14 24.34 & -17 24 06.6 & 0.990 & 4\\
SNLS-P06D4ca & ... & 22 16 16.62 & -18 02 54.7 & 0.784 & 4\\
SNLS-P06D4cb & ... & 22 15 54.01 & -17 44 08.7 & 0.999 & 4\\
SNLS-P06D4cd & SNLS-06D4fo & 22 13 49.45 & -17 36 26.2 & 0.806 & 4\\
SNLS-P06D4ce & ... & 22 17 13.68 & -17 22 24.9 & 0.969 & 3\\
SNLS-P06D4cf & SNLS-06D4fa & 22 13 39.66 & -17 24 14.6 & 0.720 & 4\\
SNLS-P06D4ch & SNLS-06D4fc & 22 13 51.06 & -17 19 31.3 & 0.678 & 4\\
SNLS-P06D4ci & SNLS-06D4hh & 22 15 52.04 & -18 13 50.5 & 0.599 & 5\\
SNLS-P06D4cj & ... & 22 13 28.59 & -18 11 57.3 & 0.187 & 5\\
SNLS-P06D4ck & SNLS-06D4gp & 22 15 53.26 & -17 47 29.8 & 0.681 & 4\\

\hline
\end{tabular}
\end{center} 
\end{table*}

\end{document}